\documentclass[manuscript]{aastex63}

\shorttitle{}
\begin{document}
\title{Radiative transfer with opacity distribution functions: Application to narrow band filters}
\author{L. S. Anusha}
\affiliation{Max-Planck-Institut f\"ur Sonnensystemforschung, Justus-von-Liebig-Weg 3, 37077 G\"ottingen, Germany}
\email{bhasari@mps.mpg.de}
\author{A. I. Shapiro}
\affiliation{Max-Planck-Institut f\"ur Sonnensystemforschung, Justus-von-Liebig-Weg 3, 37077 G\"ottingen, Germany}
\author{V. Witzke} 
\affiliation{Max-Planck-Institut f\"ur Sonnensystemforschung, Justus-von-Liebig-Weg 3, 37077 G\"ottingen, Germany}
\author{M. Cernetic} 
\affiliation{Max-Planck-Institut f\"ur Sonnensystemforschung, Justus-von-Liebig-Weg 3, 37077 G\"ottingen, Germany}
\affiliation{Max-Planck-Institut f\"ur Astrophysik, Karl-Schwarzschild-Str. 1,
D-85741 Garching, Germany}
\author{S. K. Solanki}
\affiliation{Max-Planck-Institut f\"ur Sonnensystemforschung, Justus-von-Liebig-Weg 3, 37077 G\"ottingen, Germany}
\affiliation{School of Space Research, Kyung Hee University, Yongin, Gyeonggi 446-701, Republic of Korea}
\author{L. Gizon}
\affiliation{Max-Planck-Institut f\"ur Sonnensystemforschung, Justus-von-Liebig-Weg 3, 37077 G\"ottingen, Germany}
\affiliation{Georg-August-Universit\"at, Friedrich-Hund-Platz 1, 37077 G\"ottingen, Germany}
\affiliation{Center for Space Science, NYUAD Institute, New York University Abu Dhabi, Abu Dhabi, UAE}
 
\begin{abstract}
Modelling of stellar radiative intensities in various spectral pass-bands plays an important role in stellar physics.
At the same time the direct calculations of the high-resolution spectrum and then integrating it over the given spectral pass-band is computationally demanding due to the vast number of atomic and molecular lines. This is particularly so when employing three-dimensional (3D) models of stellar atmospheres. To accelerate the calculations, one can employ approximate methods, e.g., the use of Opacity Distribution Functions (ODFs). Generally, ODFs provide a good approximation of traditional spectral synthesis i.e., computation of intensities through filters with strictly rectangular transmission function. However, their performance  strongly deteriorates when the filter transmission noticeably changes within its pass-band, which is the case for almost all filters routinely used in stellar physics. 
In this context, the aims of this paper are a) to generalize the ODFs method for calculating intensities through filters with arbitrary  transmission functions; b) to study the performance of the standard and generalized ODFs methods  for calculating intensities emergent from 3D models of stellar atmosphere. 
For this purpose we use the newly-developed  MPS-ATLAS radiative transfer code to compute  intensities emergent 3D cubes simulated with the radiative magnetohydrodynamics code MURaM. The calculations are performed in the 1.5D regime, i.e., along many parallel rays passing through the simulated cube. We demonstrate that generalized ODFs method allows accurate and fast syntheses of spectral intensities and their centre-to-limb variations.
 
\end{abstract}

\section{Introduction}
\label{intro}
Accurate modelling of stellar spectra for stars with various  effective temperatures, surface gravities, and elemental compositions is a key tool in stellar physics. In addition to these fundamental stellar parameters, the emergent spectrum is also affected by stellar magnetic activity. In particular, effects from magnetic activity might significantly complicate characterization of exoplanets and their atmospheres with transit photometry and spectroscopy. Thus, also modelling the impact of stellar surface magnetic fields on emerging stellar spectra have since recently came under the spotlight.

For a long time stellar spectra have been computed using one-dimensional (1D) atmospheric structures calculated under the assumption of radiative equilibrium \citep[with some models also including simple parameterizations of convective flux and overshooting, see, e.g.][]{Boehm_vitense_1964, Kurucz_2005_Atlas12_9, MARCS_descript2008,Phoenix_descript2013}. However, it has been shown that such 1D structures do not necessarily represent average properties of the real three-dimensional (3D) atmospheres \citep{1D_bad}. 
Furthermore, 1D modelling does not allow  self-consistently accounting for the effects of stellar surface magnetic field on the thermal structure of stellar atmospheres \citep[albeit the semi-empirical approach dates back to][but only applicable to stars with near-solar fundamental parameters]{1975SoPh...42...79S,vernazzaetal1981,1993SSRv...63....1S}.  Consequently, while semi-empirical 1D models have been successfully used for simulating the effects of surface magnetic field on the solar spectrum \citep[see][for review]{MPS_AA, TOSCA2013} their extension to modelling magnetised atmospheres of other stars is not straightforward \citep[see, e.g.,][]{Veronika2018}. 

The advent of powerful computers has made it possible to gradually replace 1D modelling with 3D hydrodynamic and magnetohydrodynamic (HD and MHD, respectively) simulations of the near-surface layers of the Sun and stars. Such simulations have reached a high level of realism \citep[see, e.g.,][]{1998ApJ...499..914S,2005A&A...429..335V,2017ApJ...834...10R,2011A&A...531A.154G,2012JCoPh.231..919F} and have been extensively tested against various solar observations. Using these simulations, the emergent spectra can then be calculated in the 1.5D regime, i.e., along many parallel rays passing through a 3D model.  The resulting intensity along any particular direction is then calculated by combining the intensities along all the rays which together sample the whole simulation cube \citep[see][for detailed descriptions of the approach]{2012A&A...547A..46H, 2013A&A...558A..20H, rietmueller-solanki-2014, norris-beck-2017}.

One of the main hurdles in calculating spectra emerging from stellar atmospheres is the intricate wavelength dependence of the opacity brought about by millions of atomic and molecular lines. These lines dominate opacity in the ultra-violet (UV) and significantly contribute to the opacity in the visible spectral domain for cool stars like the Sun. Consequently, even though many applications require computations of the spectra with intermediate spectral resolutions (e.g., 10 {\rm {\AA}} or coarser) the spectrum must still be calculated on a grid with a very fine spectral resolution to catch all the relevant spectral features (since missing them would also affect the spectrum averaged to a coarse resolution). Such calculations on a fine spectral grid are computationally expensive, especially in the case of 1.5D calculations when spectra must be synthesised, sometimes along millions of rays.

One way to drastically reduce the computational time is to use Opacity Distribution Functions (ODFs) introduced by \citet[][]{1951ZA.....29..199L}  \citep[see also][for detailed discussions on ODFs]{2014tsa..book.....H}. The ODFs method is based on a clever way of describing the opacity that significantly reduces the number of frequencies on which the radiative transfer (RT) equation is solved. In other words, the detailed frequency dependence of the opacity is replaced by a few points, which are representative of the entire opacity distribution.
ODFs have been extensively used and tested for 1D models of the solar atmosphere \citep[see e.g.,][]{1974ApJ...188L..21K,1979ApJS...40....1K, 2019A&A...627A.157C}.

Present computing resources allow direct (i.e., on a fine spectral grid) calculations of spectra with 1D models so that the ODFs method is rarely used in 1D modelling. At the same time direct calculations of emergent spectra are still not feasible over a broad spectral range for the 1.5D computations based on the output of 3D (M)HD simulations. Therefore, the transition from 1D to 3D (M)HD  modelling of stellar atmospheres has rekindled the interest in the ODFs method.

There are, however, two shortcomings of the ODFs method which have to be addressed to make it a reliable tool for comprehensive modelling of stellar spectra. First, the applicability of the ODFs method for synthesising emerging spectra from 3D (M)HD cubes has never been tested. Second, the ODFs method allows calculating radiative intensity integrated in various spectral domains, which can be considered as spectral intensity passing through a filter with a transmission function that has a rectangular dependence on the wavelength (i.e., zero outside of the given spectral pass-band and a constant value within this pass-band). However, it does not allow calculating intensity passing through filters with non-rectangular dependence of the transmission function on the wavelength. At the same time, such calculations are needed, for example, to analyse solar observations in narrow band filters \citep[e.g., performed by the Sunrise balloon borne observatory, see][]{2010ApJ...723L.127S},
to model stellar colours \citep[which are measured in different narrow and broadband filters, see, e.g.,][for a comprehensive list]{Sun_in_filters}, to calculate limb darkening coefficients in the pass-band of planetary hunting telescopes (e.g., {\it Kepler}, TESS, and CHEOPS) as well as to model the impact of magnetic activity on stellar brightness measured by various ground- and space-based filter equipped telescopes. For such calculations the spectral pass-band has to be split in many small intervals where the spectral dependence of the transmission function can be neglected. 
The large number of intervals, however, increases the number of frequencies at which spectral synthesis must be preformed and, thus, reduces the  performance of the ODFs method.

In this study we eliminate both shortcomings. Firstly, we generalise the ODFs method for calculating intensities passing through filters with arbitrary transmission functions and present the filter-ODFs (hereafter, FODFs) method. For this purpose we limit ourselves to filters that are narrow enough that neither continuum opacity nor Planck function changes noticeably within the filter. Secondly we test the performance of the ODFs and FODFs methods on the exemplary case of solar 3D MHD simulations with the MURaM code \citep[Max Planck Institute/University of Chicago Radiative MHD, see][]{2005A&A...429..335V}. For the RT calculations we use the newly developed {\bf{M}}erged {\bf{P}}arallelised {\bf{S}}implified (MPS) - ATLAS code \citep{2020A&A_Witzke}.  A preliminary study of the filter-ODFs method was performed by \citet[][]{2019A&A...627A.157C}, who demonstrated that the method is capable of accurately returning intensity in the Str\"{o}mgren {\it b} filter in the  case of 1D semi-empirical models of the solar atmosphere \citep[FALC and FALP by][]{1993ApJ...406..319F}. Here we present a more detailed study based on 3D MHD simulations.

The paper is organized as follows. Section~\ref{method} gives a general overview of the ODFs method and presents the FODFs method, Section~\ref{results} presents the main
results of the paper, and Section~\ref{conclusions} summarises the conclusions where we discuss the applications and limitations of the ODFs/FODFs method.

\section{Methods}
\label{method}
\subsection{The (Filter) Opacity Distribution Functions}
\label{fodfs-method}
In this section we first recall the traditional ODFs method used for calculating radiative intensity passing through rectangular filters. Then we introduce the formalism of the FODFs method that we developed for calculating radiative intensities passing through an arbitrary filter.

Let us consider a wavelength interval $ [\lambda_1,\lambda_2] $ over which we would like to compute the total spectral intensity:
\begin{equation}
\label{eq:f}
f=\int_{\lambda_{1}}^{\lambda_{2}} I(\lambda) d\lambda.   
\end{equation}
The most straight forward way of computing the integral in Equation~(\ref{eq:f}) is to calculate the emergent intensity on a fine grid of spectral points within the $ [\lambda_1,\lambda_2] $ interval (hereafter, spectral bin) and approximate $f$ by the quadrature $f_{\rm HR}$ (where ``HR'' stands for high resolution since the quadrature must be calculated on the high-resolution wavelength grid):
\begin{equation}
\label{eq:f_quadr}
f \approx f_{\rm HR} \equiv \sum\limits_{i=1}^{N} I(\lambda_i) \cdot \Delta \lambda,
\end{equation}
where $N$ is the number of points in a  high resolution grid, which we assume to be equidistant, and $\Delta \lambda = (\lambda_2-\lambda_1) / (N-1)$ is one step on the grid.

Equation~(\ref{eq:f_quadr}) implies that, firstly, the opacities in a 1D atmosphere (or along a ray following the 1.5D approach) must be calculated at $N$ frequency points and, secondly, the formal solution of the RT equation must be performed $N$ times.  A great simplification can be achieved by employing the Local-thermodynamic-equilibrium (LTE) approximation. Then the opacity at a given frequency only depends on the elemental composition, pressure, temperature, and turbulent velocity. 
We note that here we approximate the detailed velocity profile by a scalar generally known as the micro-turbulence parameter which broadens the lines and enters opacity. Such an approximation does not allow one to accurately synthesise profiles of individual spectral lines but it is routinely applied for calculating intensities in various pass-bands \citep[see, e.g.,][]{norris-beck-2017, Yelles2020}.
Consequently, for a given elemental composition  the opacity can be pre-tabulated and then interpolated for all the points in the stellar atmosphere. This avoids direct calculations of the opacity values on the fly and, dramatically reduces the computational cost. 

Nevertheless, calculations with the quadrature method, especially over the broad frequency interval, can be very time-consuming. The main problem is that $N$ often must be a very large number to find a reasonable approximation of the integral $f$, so that one has to solve the RT equation many times. We illustrate this point in Figure~\ref{figure1} whose top panels show high-resolution opacity calculated with the MPS-ATLAS code (see below) under the condition of LTE for the temperature 4250 K and pressure 3.3$\times 10^4$ dyn/cm$^2$ (values close to those that can be met in the upper solar photosphere).  One can see that 
the opacity is fully dominated by the Fraunhofer lines in the UV spectral domain (left top panel of Figure~\ref{figure1}) and is strongly affected by lines in the visible (right top panel of Figure~\ref{figure1}). As a result the opacity has a highly complex spectral profile and can vary over many orders of magnitude within a very narrow spectral interval. The emergent spectra are plotted in the bottom panels of Figure~\ref{figure1}. They are calculated for one of the rays passing through the MHD cube representing the solar atmosphere (see below for more details). One can see that just like the  opacity the spectra have a very rich structure. In particular, the numerous absorption lines in the UV so strongly overlap and blend each other that it even becomes impossible to identify the continuum level. All in all, Figure~\ref{figure1} illustrates that one needs a very fine frequency grid (i.e., up to a thousand points) to catch all spectral details shown in middle and bottom panels of Figure~\ref{figure1} to accurately calculate the integrated intensity in the shown spectral domains.

An alternative and much more time efficient approach compared to direct calculation of the quadrature $f_{\rm HR}$ is the ODFs method. This method sorts the frequency points in the high-resolution grid according to the value of opacity at some depth point in the atmosphere, i.e., to come up with a transformation $i \rightarrow M (i)$, which implies that the $i$-th grid point (with $1<i<N$) is replaced with the $M(i)$-th point (with $1<M(i)<N$). The main assumption of the ODFs method is that the transformation $M$ is the same for all depth points in the atmosphere contributing to the intensity in the bin $ [\lambda_1,\lambda_2] $.  Under such an assumption Equation~(\ref{eq:f_quadr}) can be rewritten as
\begin{equation}
\label{eq:f_quadr_sort}
f_{\rm HR} = \sum\limits_{i=1}^{N} I(\lambda_{M(i)}) \cdot \Delta \lambda.
\end{equation}
There is a crucial difference between Equation~(\ref{eq:f_quadr}) and Equation~(\ref{eq:f_quadr_sort}) introduced by reordering terms in the summation. 
Indeed, in Equation~(\ref{eq:f_quadr}) the term with index $i$ and those with indices close to $i$ can correspond to very different values of opacity and, consequently, be very different.
In contrast, in Equation~(\ref{eq:f_quadr_sort}) nearby terms, e.g., with indices $M(i)$,
$M(i-1)$, and $M(i+1)$, 
should have similar values of opacity (according to the ODFs assumption that all depths points contribute to the intensity  in the bin [$\lambda_1$, $\lambda_2$]) and emergent intensity. 
Let us then rewrite Equation~(\ref{eq:f_quadr_sort}) splitting the entire sum into $s$ sub-sums with the $j$-th sub-sum containing $N_j$ grid points:
\begin{equation}
\label{eq:f_group}
f_{\rm HR} = \left ( \sum\limits_{i=1}^{N_1} I(\lambda_{M(i)})  + 
\sum\limits_{i=N_1+1}^{N_1+N_2} I(\lambda_{M(i)})   + 
\sum\limits_{i=N_1+N_2+1}^{N_1+N_2+N_3} I(\lambda_{M(i)})   +...+
\sum\limits_{i=N-N_s+1}^{N} I(\lambda_{M(i)}) \right ) \cdot \Delta \lambda .
\end{equation}
The opacity and intensity values for all points within individual sub-sums are close to each other. Thus, one can now substitute the intensity values in the sub-sums by the intensity calculated using opacity averaged over all the frequency points in the sub-sum, i.e.,
\begin{equation}
\label{eq:subbin}
f_{\rm HR} \approx  f_{\rm ODFs} \equiv \sum\limits_{j=1}^{s} \widehat{I_j} \, \widehat{\Delta \lambda_j},
\end{equation}
where $\widehat{I_j}$ is calculated by solving the RT equation with opacity averaged (e.g., arithmetically or geometrically, see Section~\ref{results}) over the grid points within the sub-sum, and $\widehat{\Delta \lambda_j} = N_j \cdot \Delta \lambda$. The $s$ intervals are called sub-bins and the mean opacity in each of the sub-bins is called the ODF. Exactly as the high-resolution opacity values, the ODFs for a given chemical composition can be pre-tabulated as a function of temperature, pressure, and micro-turbulent velocity. Here $\widehat{I_j}$ is formulated to represent an Intensity Distribution Function associated with the ODF.

The first step towards moving from the ODFs to FODFs is a generalisation of Equations~(\ref{eq:f_group}--\ref{eq:subbin}) to a non-equidistant high-resolution spectral grid. Equation~(\ref{eq:f_group}) then becomes:

\begin{equation}
\label{eq:f_group_non}
f_{\rm HR} =  \sum\limits_{i=1}^{N_1} I_{M(i)} \Delta \lambda_{M(i)}
+ 
\sum\limits_{i=N_1+1}^{N_1+N_2} I_{M(i)} \Delta \lambda_{M(i)}   + 
\sum\limits_{i=N_1+N_2+1}^{N_1+N_2+N_3} I_{M(i)} \Delta \lambda_{M(i)}  +...+
\sum\limits_{i=N-N_s+1}^{N} I_{M(i)} \Delta \lambda_{M(i)} ,
\end{equation} where for simplicity we introduce a designation $I_{M(i)} \equiv  I(\lambda_{M(i)})  $.

Each sub-sum in Equation~(\ref{eq:f_group_non}) can be written as $\bar{I}_j \times l_j$, a product of weighted mean of intensity (with weights being proportional to the steps of the wavelength grid) $\bar{I}_j=\sum I_{M(i)} \left ( \Delta \lambda_{M(i)} / l_j \right )$, and the width of the sub-bin $l_j =  \sum \Delta \lambda_{M(i)}$. Here we also define the size of the $j$-th sub-bin as
\begin{equation}
    \label{sub-bin-sizes}
    s_j=l_j/(\lambda_2-\lambda_1).
\end{equation}
Then, exactly like in Equation~(\ref{eq:subbin}) the weighted mean of intensity $\bar{I}_j$ can be approximated by the intensity calculated with the weighted mean of opacity denoted as $\widehat{I}_j$. Most importantly, the utilised weights only depend on the high-resolution frequency grid, so that the weighted mean opacity in each of the sub-bins can be pre-calculated and pre-tabulated, and we now call them FODFs. Similar to the case of ODFs, here $\widehat{I_j}$ is formulated to represent an Intensity Distribution Function associated with the FODF. Hereafter for convenience these Intensity Distribution Functions are simply referred to as intensities computed using ODFs or FODFs.

All in all, the ODFs method implies approximation of the mean intensity in the sub-bin by intensity calculated with the mean opacity in the sub-bin. The accuracy of such an approximation strongly depends on the choice of the splitting in Equation~(\ref{eq:f_group}), which is a crucial aspect of the ODFs method. A proper choice implies that the intensity in the sub-bin should depend on the opacity in a roughly linear way. In particular, one should avoid situations when intensity strongly depends on the opacity in some part of the sub-bin but then saturates in another.  As a result, the relative sizes of the sub-bins (defined in Equation~(\ref{sub-bin-sizes})) with large opacity values are usually chosen to be smaller than those with small opacity values. A detailed study of the optimal choice of the sub-bins was recently performed by \cite{2019A&A...627A.157C}.

Let us now consider the case when instead of calculating the total spectral intensity in the bin (or in other words, intensity passing through the rectangular filter) one needs to calculate intensity passing trough 
a filter with some wavelength-dependant transparency $\phi=\phi(\lambda)$ (with $0 \le \phi \le 1$ ):
\begin{equation}
\label{eq:F}
F=\int_{\lambda_{1}}^{\lambda_{2}} I(\lambda) \, \phi(\lambda) \, d\lambda.   
\end{equation}
Similar to Equation~(\ref{eq:f_quadr}) the
integral might be approximated by the quadrature:
\begin{equation}
\label{eq:F_quadr}
F \approx F_{\rm HR} \equiv \sum\limits_{i=1}^{N} I(\lambda_i) \, \phi(\lambda_i) \cdot \Delta \lambda,
\end{equation}

The most direct way of employing the ODFs method for calculating $F_{\rm HR}$ is to move the filter transmission function out of the summation in Equation~(\ref{eq:F_quadr}) and write the intensity as a product of the mean transmission value and intensity in the rectangular filter (which can then be approximated by $f_{\rm ODFs}$):
\begin{equation}
\label{eq:F_ODF}
F_{\rm HR}  \approx F_{\rm ODFs} \equiv \left (\frac{1}{N}  \sum\limits_{i=1}^{N}  \phi(\lambda_i) \right ) \cdot f_{\rm ODFs},
\end{equation}
where $f_{\rm ODFs}$  is given by Equation~(\ref{eq:subbin}).  An obvious advantage of such an approach is that it does not require any recalculations of ODFs, so that the same ODFs can be used for all filters. It is, however, clear that $F_{\rm ODFs}$ cannot be an accurate approximation of $F_{\rm HR} $ since it does not account for the distribution of spectral features within the filter. 

A much more accurate way is to sort and split the terms in Equation~(\ref{eq:F_ODF}) exactly as done in Equations~(\ref{eq:f_group}) and (\ref{eq:f_group_non}). Then, in each of the sub-bins one can approximate the mean values of intensity weighted with the transmission function by intensity calculated with the corresponding  weighted mean of opacity. The only disadvantage of such an approach is that the transmission function is bound to affect the optimal choice of the sub-binning, i.e., grouping of terms in Equations~(\ref{eq:f_group}) and (\ref{eq:f_group_non}). For example, it might happen that the largest opacity coincidentally corresponds to small values of the transmission function (i.e., a strong spectral line is located in the wings of the filter). Then it does not make sense to approximate the peak of the opacity distribution (i.e., highest values of the opacity) as accurately as one would do it in the case of the rectangular transmission function.

A simple way to overcome this problem is to transform a high-resolution 
wavelength grid $\lambda_i$:
\begin{equation}
\label{eq:tilda}
\tilde{\lambda}_i=\Delta \lambda \, \sum\limits_{j=1}^{i}  \phi(\lambda_j), \,\,\,\,\,\, \Delta  \tilde{\lambda}_i = \Delta \lambda \, \phi(\lambda_i),
\end{equation}
where the transformation is for simplicity written for the case of an equidistant $\lambda_i$ grid.

Then Equation~(\ref{eq:F_quadr}) can be rewritten as
\begin{equation}
\label{eq:F_quadr_tilde}
F_{\rm HR} = \sum\limits_{i=1}^{N} I_i    \,  \Delta \tilde{\lambda}_i.
\end{equation}
Equation~(\ref{eq:F_quadr_tilde}) represents a quadrature with non-equidistant grid which can be adequately approximated by the ODFs sum similar to that in Equation~(\ref{eq:subbin}), but
written on the new $\tilde{\lambda}$ wavelength grid:
\begin{equation}
    F_{\rm FODFs} \equiv \sum\limits_{j=1}^{s}  \widehat{I_j}  \,\widehat{ \Delta \tilde{\lambda_j}},
\end{equation}
where $F_{\rm FODFs}$ stands for intensity calculated with filter-ODFs (which is basically normal ODFs, but calculated on the new wavelength grid $\tilde{\lambda}$),  $\widehat{I_j}$ is intensity computed with opacity averaged over the sub-bin with width $\widehat{ \Delta \tilde{\lambda_j}}$.
The most important benefit of transforming the 
wavelength grid is that sizes of sub-bins ($s_j$ defined in Equation~(\ref{sub-bin-sizes})) can be taken the same as for the rectangular filter, e.g., one can directly use receipts proposed by \cite{2019A&A...627A.157C}. Furthermore,  FODFs can be pre-calculated and tabulated for any given filter profile as a function of temperature, pressure and micro-turbulent velocity values exactly the same way the traditional ODFs are pre-tabulated.

We note here that dividing a given bin into sub-bins with equal or unequal widths is respectively termed uniform or non-uniform sub-binning.
For example, in the earlier formulations Kurucz used twelve non-uniform sub-bins per bin, such that the sub-bins containing largest values of the opacity are closely spaced \citep[see, e.g.,][]{1991ASIC..341..441K,Castelli_2005_DFSYNTHE}. Hereafter we refer to this particular sub-binning suggested by Kurucz, as the Kurucz-sub-binning (see also Table~\ref{tab:table2} and discussions in Section~\ref{results}).
The Kurucz sub-binning set-up of twelve non-uniform sub-bins per bin is used for most of our tests.

We stress here that the ODFs method is based on the assumption that transformation $i \rightarrow M (i)$ (see Equation~(\ref{eq:f_quadr_sort})) does not change within the region where radiation in a given bin forms. In other words, we assume that there is a unique correspondence between a given point in a sub-bin and the actual wavelength for all depths within the formation region. For instance, the wavelength of the point of maximum opacity in the given bin is the same at all depths. 
The accuracy of the ODFs method diminishes if this assumption is not fulfilled, e.g., when maximum value of opacity corresponds to a certain wavelength for some depths, but to another wavelengths elsewhere. In Section~\ref{results} we compare the intensities computed using ODFs method and those computed using the high-resolution opacities and, thus, test the ODFs assumption.

\begin{table}[h!]
  \begin{center}
    \caption{Mean errors of $e_{\textrm {ODFs}}$ plotted in Figures~\ref{figure2} and \ref{figure3}.}
    \label{tab:table1}
    \begin{tabular}{c|c|c} 
      \textbf{UV: High-resolution vs.} & \textbf{UV: High-resolution vs.} & \textbf{UV: High-resolution vs.} \\
      \textbf{ODF-1 bin} & \textbf{ODF-5 bin} & \textbf{ODF-20 bin} \\
      \textrm{GM} & \textrm{GM} & \textrm{GM} \\
      \hline
      \hline
       -1.22 \% & -1.07 \% &  -1.14 \% \\
      \hline
      \hline
      \textrm{AM} & \textrm{AM} & \textrm{AM} \\
      \hline
      \hline
       -0.68 \% & -0.53 \% & -0.65  \% \\
      \hline
      \hline
      \textbf{Visible: High-resolution vs.} & \textbf{Visible: High-resolution vs.} & \\
            \textbf{ODF-1} & \textbf{ODF-5} & \\
      \textrm{GM} & \textrm{GM} &  \\
      \hline
      \hline
        -0.18 \% &  -0.19 \% & \\
      \hline
      \hline
      \textrm{AM} & \textrm{AM} &  \\
      \hline
      \hline
        0.19 \% &  0.14 \% & \\
      \hline
      \hline
    \end{tabular}
  \end{center}
\end{table}

\begin{table}[h!]
  \begin{center}
    \caption{Sub-bin sizes $s_j$ (defined in Equation~(\ref{sub-bin-sizes})) used for the plots in Figure~\ref{figure4}.}
    \label{tab:table2}
    \begin{tabular}{l|c} 
      \textbf{Figure legend} & \textbf{Sub-bin sizes $s_j$}  \\
      \hline
      \hline
       nonuni-12 & \{0.1, 0.2, 0.3, 0.4, 0.5, 0.6, 0.7, 0.8, 0.9, 0.95, 0.9833, 1.0\} \\
       uni-12 &  \{$8.3\times10^{-2}$, 0.166, 0.25, 0.333, 0.416, 0.5, 0.583, 0.666, 0.75, 0.837, 0.916, 1.0\}\\
       uni-8 & \{0.125, 0.25, 0.375, 0.5, 0.625, 0.75, 0.875, 1.0\}\\
       uni-4 & \{0.25, 0.5, 0.75, 1.0\}\\
       nonuni-4a & \{0.32, 0.6, 0.8, 1.0\} \\
       nonuni-4b & \{0.80, 0.92, 0.96, 1.0\}\\
       nonuni-4c & \{0.76, 0.88, 0.94, 1.0\}\\
      \hline
      \hline
    \end{tabular}
  \end{center}
\end{table}
\begin{figure}
\includegraphics[scale=0.4]{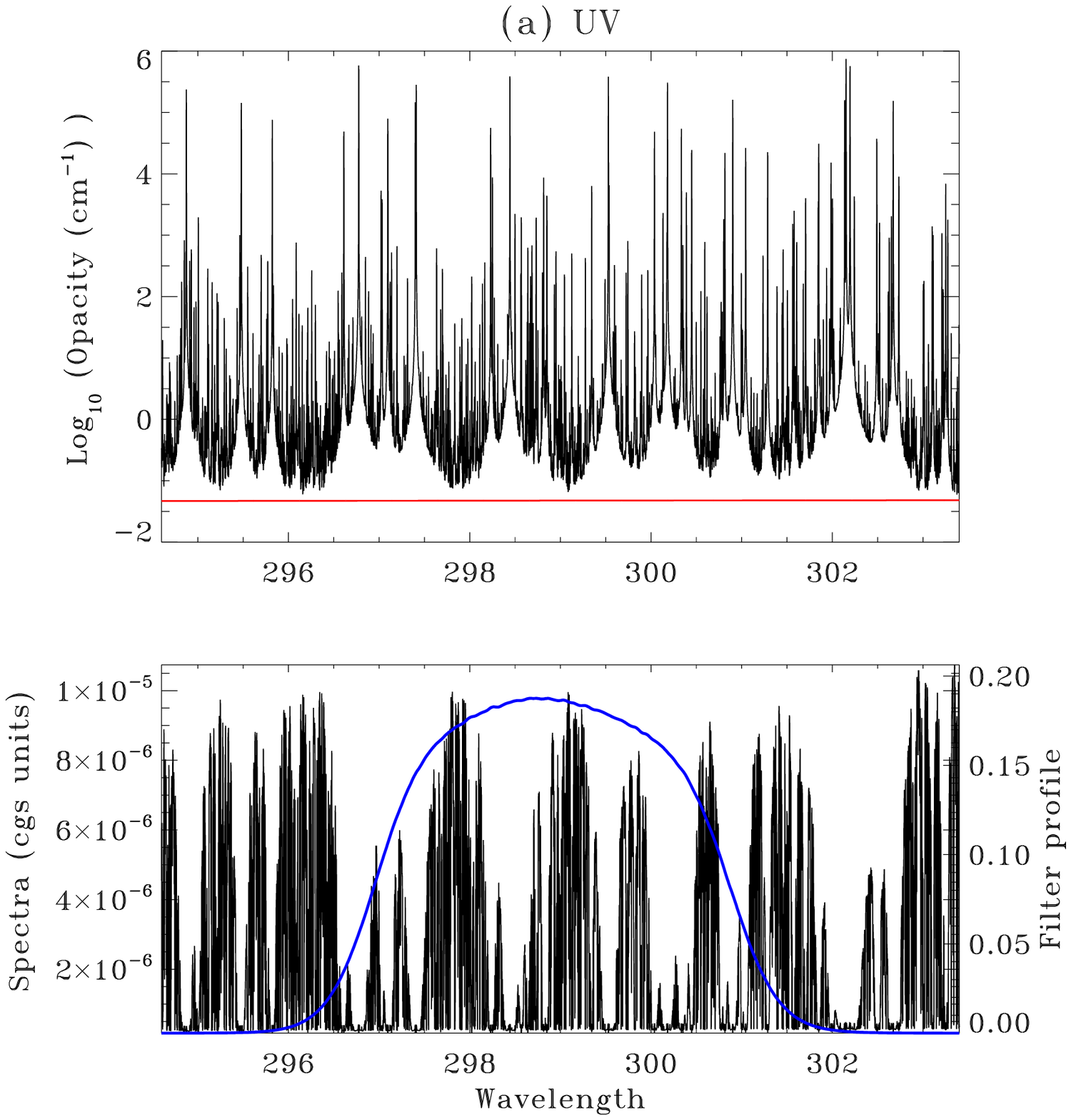}
\includegraphics[scale=0.4]{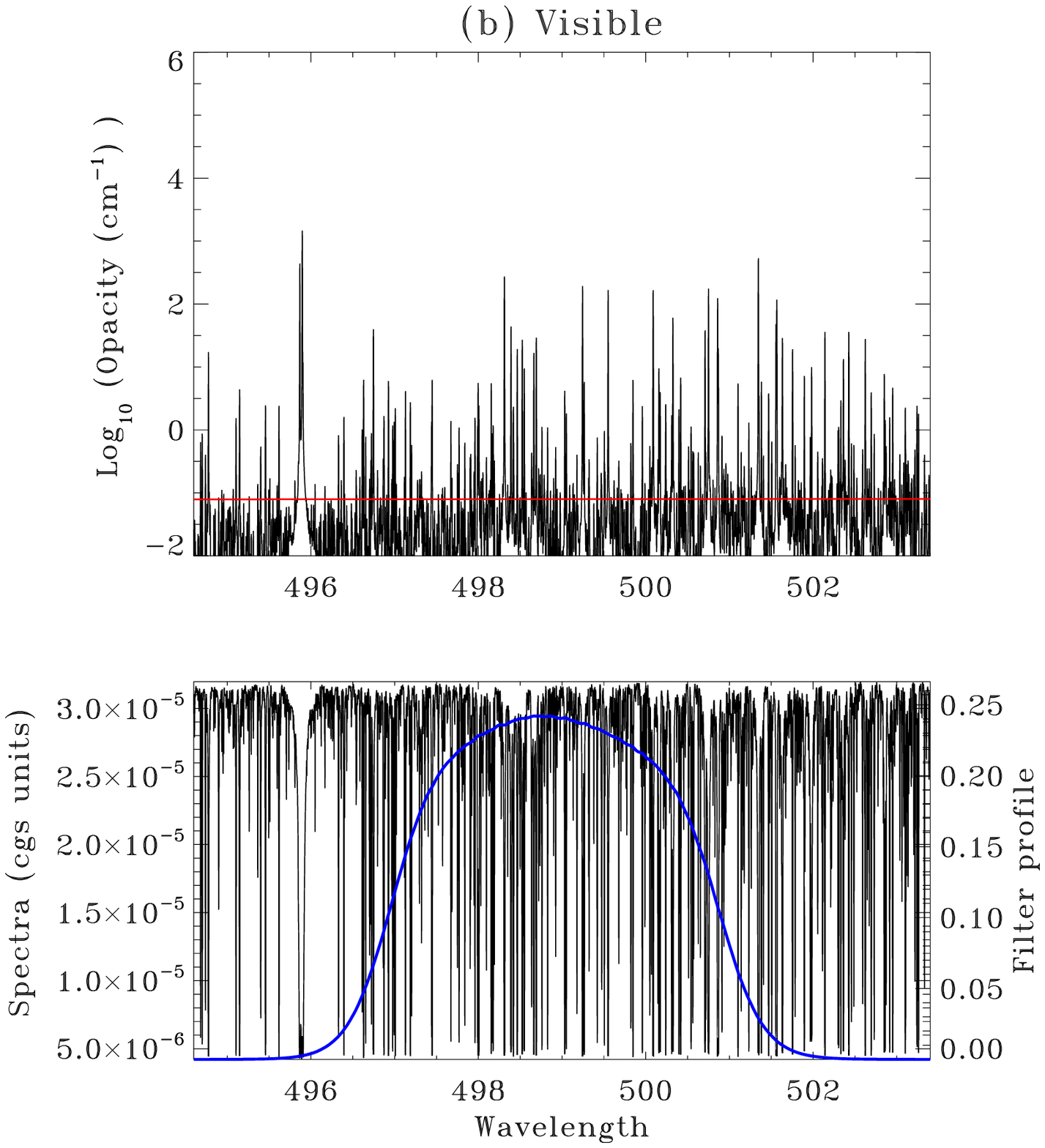}
\caption{Top panels show high resolution opacities (black lines) in the UV and the visible pass-band over-plotted
with the continuum opacity (red lines) at a given depth point in the line formation region. 
Bottom panels show high resolution emergent spectra (black lines). Also over-plotted are 
the SuFI and SuFIV filter profiles (blue lines) in the respective spectral pass-bands.
}
\label{figure1}
\end{figure}

\subsection{Numerical setup}
\label{1.5D-RT}
The 3D atmospheric structures utilised for tests in this paper were computed using the radiative MHD code MURaM \citep[][] {2005A&A...429..335V}. The MURaM code was designed for realistic simulations of the stellar upper convection zones and photospheric layers. To this end, the non-ideal MHD equations are numerically solved in a local Cartesian box.
The code includes a non-grey treatment of the energy exchange between matter and radiation, namely the RT is solved independently in four opacity bins \citep[see][for a detailed description]{2005A&A...429..335V}.

In this paper we used the 3D MHD atmospheric structures of \citet[][]{2020ApJ...894..140R} who used the same simulation set-up as in \citet[][]{Rempel2014}. 
These simulations use vertical or symmetric lower boundary and a variety of initial magnetic field conditions to enable small-scale dynamo action and obtain various levels of magnetization \citep[see e.g.,][]{2020ApJ...894..140R}. For our studies we chose the case with an initial 300 G vertical magnetic field, that corresponds to the most magnetized cube, in order to cover a broader variety of structures and, thus, provides a better testbed for ODFs and FODFs. 

The MHD cubes that we used for our study cover only $1$ Mm above the photosphere. 
Although the formation heights of most of the lines  lie within the cube, it is not sufficiently high to cover the formation region of some of the UV lines in the spectral region of our interest (see discussion at the end of this section). 
Such a poor sampling of the formation region will affect high-resolution calculations and F(ODF)s differently, thus leading to deviations between F(ODF)s and high-resolution calculations which are not introduced by the (F)ODFs approximation.
Therefore, to make sure that our cubes are sufficiently high we extrapolate them upward. We add twenty extra height points above the top surface of the simulation box. At these points, we keep the temperature fixed (to the value of temperature at the top most grid point of the cube before extrapolation), and assign a monotonically outward-decreasing column mass and pressure.
This ad-hoc extrapolation is justifiable since the goal of this paper is not to make a realistic calculations but rather test the (F)ODFs method by comparing it with high resolution computations on the same atmosphere. This extrapolation is used for all the UV results presented in this paper.

We follow the 1.5D approach for calculating the emergent spectrum, i.e., we treat the 1D rays piercing the 3D MHD cube as independent 1D atmospheres. For the vertical rays the temperature, density, pressure and velocity  are taken directly from the MURaM grid points, while for the inclined rays these quantities are interpolated. 

To solve the RT equation along the ray and to synthesise ODFs and FODFs we employ the newly developed MPS-ATLAS code \citep{2020A&A_Witzke}, which  combines an LTE RT solver, with a tool to calculate and pre-tabulate line opacity as well as to create ODFs and FODFs out of them. The MPS-ATLAS builds on the ATLAS9  \citep{Kurucz_manual_1970,Kurucz_2005_Atlas12_9,Castelli_ATLAS12_2005} and the DFSYNTHE  \citep{Castelli_2005_DFSYNTHE} codes. For the line opacity the standard Kurucz linelist\footnote{http://kurucz.harvard.edu/linelists/linescd/} is used, and line profiles are described by the Voigt function, except for hydrogen lines  which are treated using Stark profiles \citep{SYNTHE_2002_Cowley}. 
For solving the RT equation MPS-ATLAS first calculates the equilibrium number densities in LTE for a large set of elements, and molecules. In the next step the continuous opacity is computed, which includes all relevant opacity sources, i.e., free-free and bound-free  atomic and molecular transitions, as well as   
electron scattering and Rayleigh scattering. Subsequently, line opacity from the pre-tabulated opacity table is read, interpolated, and added. Finally, by using a Feautrier method the RT equation is solved. This is repeated for every frequency point for which the emergent intensity is needed. For more details see \citet{2020A&A_Witzke}.

For applications in solar physics the UV regime is important, at the same time the visible regime is interesting for stellar physicists. In order to represent both the regimes we choose a 10 nm wide spectral pass-band in the UV (294.6 nm-303.4 nm) and in the visible (494.6 nm-503.4 nm). The UV filter profile corresponds to one of the filters in the Sunrise Filter Imager \citep[SuFI, see][]{2011SoPh..268...35G}, while the visible filter is similar to the SuFI filter in profile shape and width, but shifted in wavelength to the blue-green part of the visible solar spectrum. Hereafter we refer to these filter profiles as SuFI and SuFIV for convenience. All the tests performed in this paper use
these exemplary filter profiles and they are shown in Figure~\ref{figure1}.

\section{Results and discussions}
\label{results}
In this section we present the results from the numerical studies that we carried out in order to test the performance of the ODFs and FODFs methods.
Namely, we study the distribution of the errors in the spectral intensities computed using ODFs and FODFs relative to that computed using the high-resolution opacity. For this purpose we used the high-resolution wavelength grid with a resolving power of R=500000.
For the convenience of the discussions we define these relative errors as (see Sect.~\ref{fodfs-method} for definitions of the terms in the right hand sides of the equations below)
$$
e_{\textrm {ODFs}}=(f_{\textrm {HR}}-f_{\textrm {ODFs}})
  /f_{\textrm {HR}}\times 100,
$$
that quantifies the accuracy of the ODFs in the rectangular filters,
$$
E_{\textrm {ODFs}}=(F_{\textrm {HR}}-F_{\textrm {ODFs}})/F_{\textrm {HR}}\times 100,
$$
that quantifies the accuracy of the ODFs in the non-rectangular filters, and
$$
E_{\textrm {FODFs}}=(F_{\textrm {HR}}-F_{\textrm {FODFs}})/F_{\textrm {HR}}\times 100,
$$
that quantifies the accuracy of the FODFs. 
We calculate all the three quantities for each of the vertical rays piercing the 300 G MHD cube from \citet{2020ApJ...894..140R} and analyze the resulting distributions of the errors. 
We note
that we utilise the same approximation of actual velocities by the micro-turbulence for calculating intensities on the high-resolution spectral grid  ($F_{\textrm {HR}}$ and  $f_{\textrm {HR}}$, see  Equations~(\ref{eq:f_quadr})~and~(\ref{eq:F_quadr})) as the one used for calculating intensities with the ODFs and FODFs approaches (i.e., $F_{\textrm {ODFs}}$ and $F_{\textrm {FODFs}}$). We use a micro-turbulent velocity of 0 km/s for the calculations presented in this paper{\bf{, unless specified otherwise}}.

\begin{figure}
\includegraphics[scale=0.4]{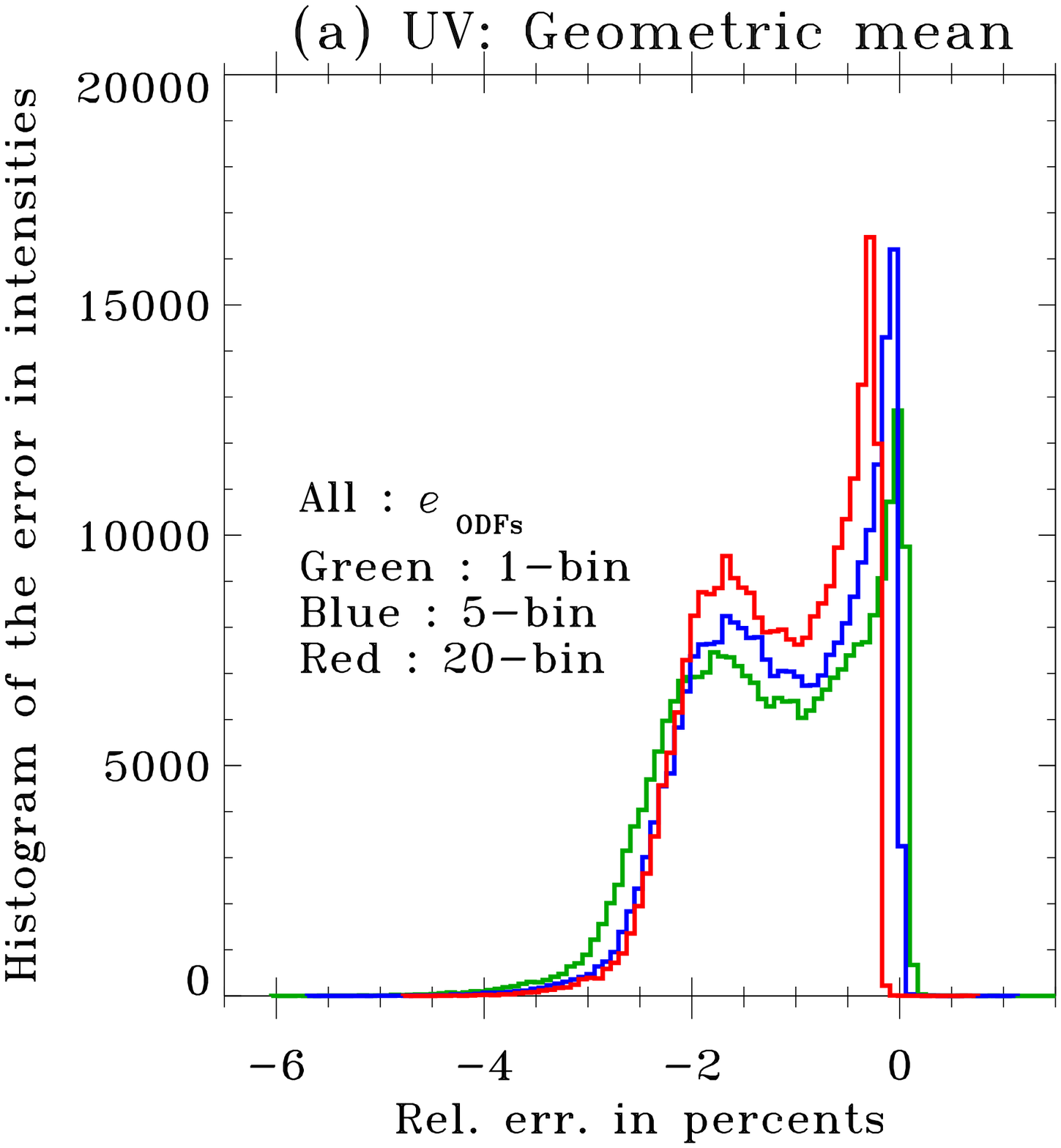}
\includegraphics[scale=0.4]{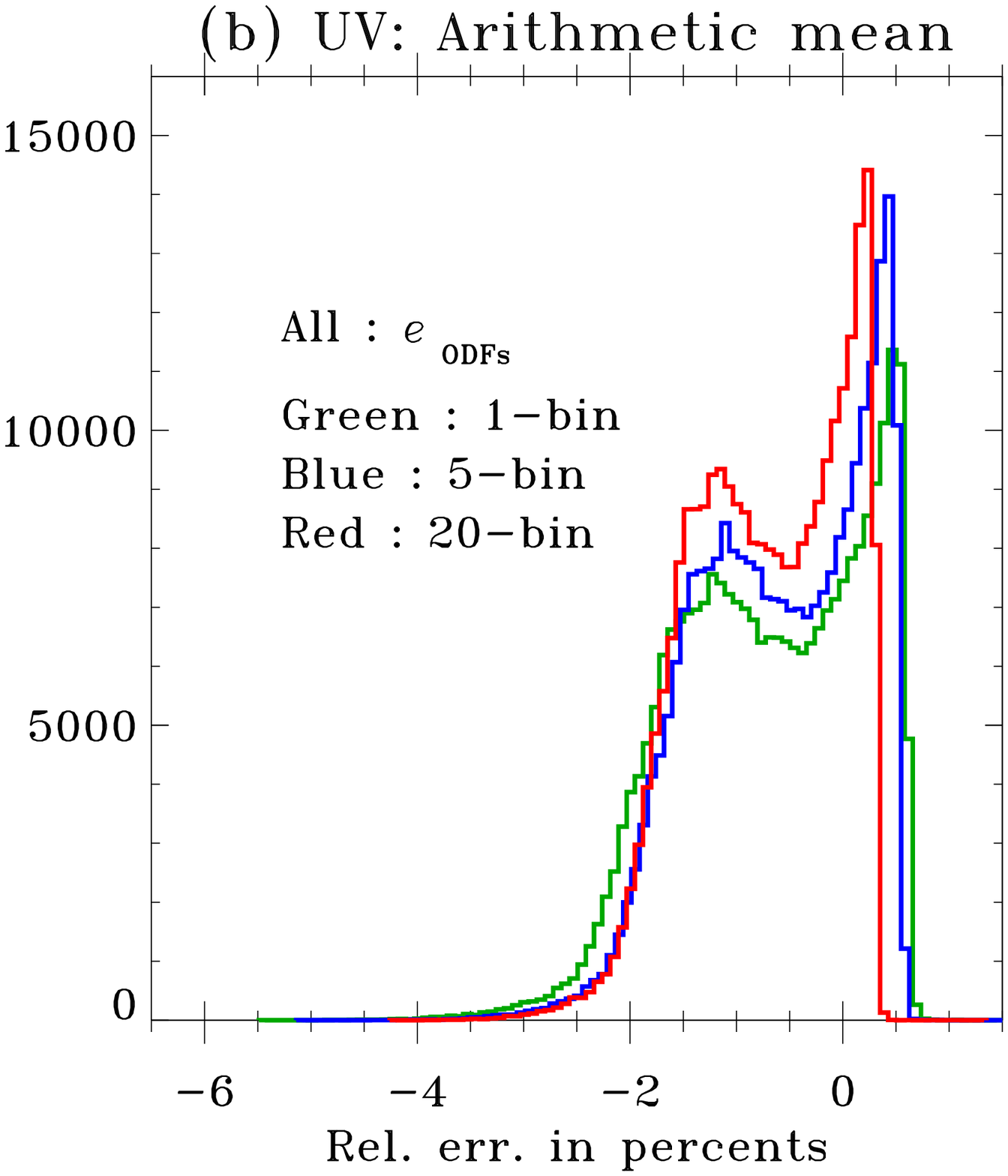}
\caption{Effect of binning.
Plotted are the histograms of $e_{\textrm {ODFs}}$ with intensities computed using one, five and twenty bins in the UV pass-band (294.6 nm-303.4 nm). The Kurucz sub-binning set-up is used. Panels: (a) Histograms of errors introduced by using the GM for computing the ODFs.
(b) The same as for panel (a), but now using the AM for computing the ODFs. 
}
\label{figure2}
\end{figure}

\begin{figure}
\includegraphics[scale=0.4]{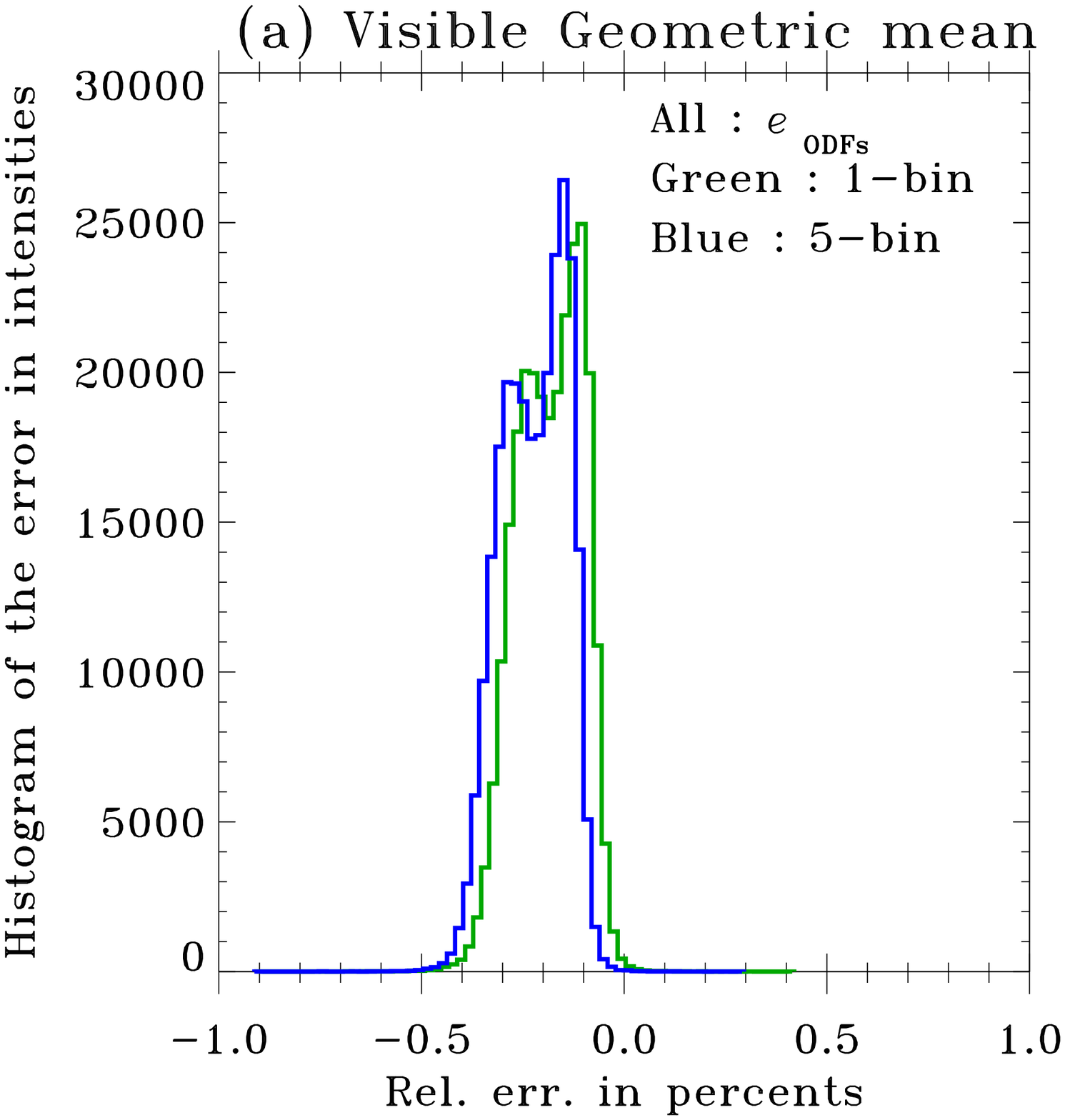}
\includegraphics[scale=0.4]{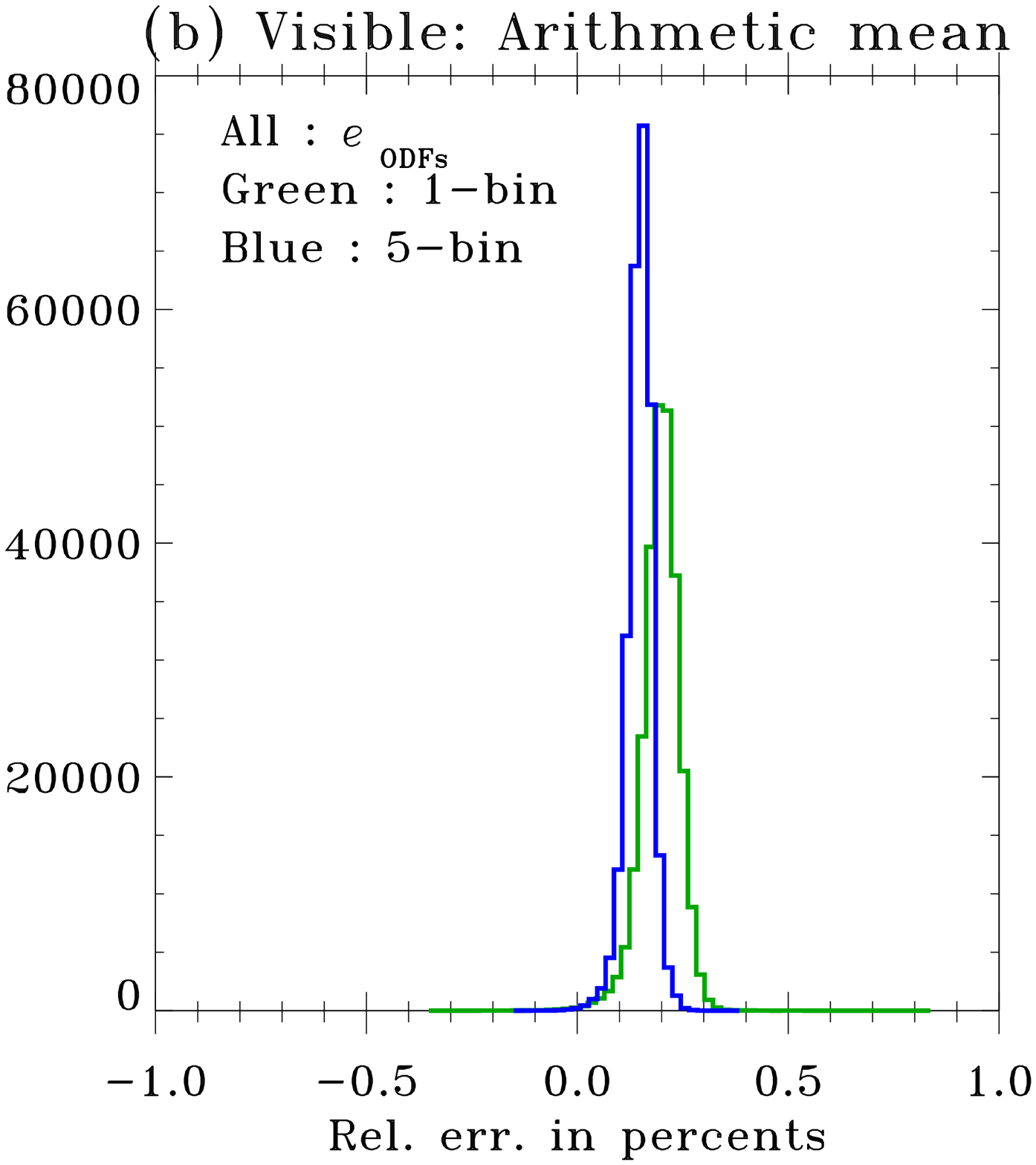}
\caption{The same as Figure~\ref{figure2}, but now for the visible pass-band (494.6 nm-503.4 nm). Note that only one and five bin ODFs are considered for this pass-band.
}
\label{figure3}
\end{figure}

\subsection{Effect of binning}
\label{results:binning} 
In this section we study $e_{\textrm {ODFs}}$ for various cases defined by the number of bins per filter pass-band, to understand how the number of bins affect the performance of the ODFs.  In the original formulation the DFSYNTHE code uses the geometric mean (GM) for averaging the opacity in the sub-bins. To study the effect of different types of the means we test both the GM and the arithmetic mean (AM) here. 

In Figures~\ref{figure2} and Figure~\ref{figure3} we show the distribution of $e_{\textrm {ODFs}}$ in the UV and the visible pass-bands, respectively. The calculations are performed with the Kurucz (non-uniform twelve) sub-binning set-up. We note that for the calculation of the ODFs we need to have a sufficient number of the high-resolution grid points in each of the sub-bins. While the wavelength grid in the DFSYNTHE code is sufficiently fine for splitting the UV pass-band in twenty bins, that is not the case for the visible pass-band. Therefore we do not consider the twenty-bin case in the visible pass-band.

In Table~\ref{tab:table1} we list the mean errors in each of the cases plotted in Figures~\ref{figure2}
and \ref{figure3}. We find that the maximum of the absolute mean error of the ODFs method is 1.22 \% in the UV and 0.19 \% in the visible pass-band relative to the  high-spectral resolution calculations. Further, we find that in the UV, the absolute mean errors are smaller for the AM than for the GM, while in the visible case they are comparable. 

The averaged opacity values computed using the AM are always larger than or equal to those computed using the GM. 
Larger opacity values imply that the corresponding radiation is formed in higher layers of the atmosphere. Since the temperature decreases with height for most of the rays we consider, the AM produces lower intensities than the GM. 
Therefore, in both the figures we see that the distributions that use the AM are shifted to the 
positive side of the horizontal axis. 
In order to represent both types of the means hereafter we use the GM for the UV pass-band and the AM for the visible pass-band.

A crucial observation one can make here is that the calculations with one bin are basically as accurate as those with a larger number of bins for both
the UV and the visible pass-bands.
In other words, the intensity computed using 240 (60) frequencies in the UV (visible) is as good as the intensity with twelve frequencies in a spectral pass-band of $\sim$ 10 nm. Thus, we conclude that in narrow spectral pass-bands,
it is sufficient to consider just one bin to reach the intrinsic accuracy of the ODFs method. 
Hereafter we continue our studies with one bin per filter unless stated otherwise. The effect of binning on broader spectral pass-bands (i.e., on those where either the Planck function or the continuum opacity substantially changes within the filter) is reserved for future studies.

\subsection{Effect of sub-binning}
\label{results:subbinning} 
We now proceed to study $e_{\textrm {ODFs}}$ for various sub-bin configurations to understand the effect of sub-binning. 
We note that our goal is not to find an optimal sub-bin configuration as was done by \citet[][]{2019A&A...627A.157C} for 1D spectral synthesis.  
Instead we only study a few examples 
to demonstrate the effect the sub-binning has on the calculations of the spectral intensities emerging from 3D atmospheres. 

\begin{figure}
\includegraphics[scale=0.4]{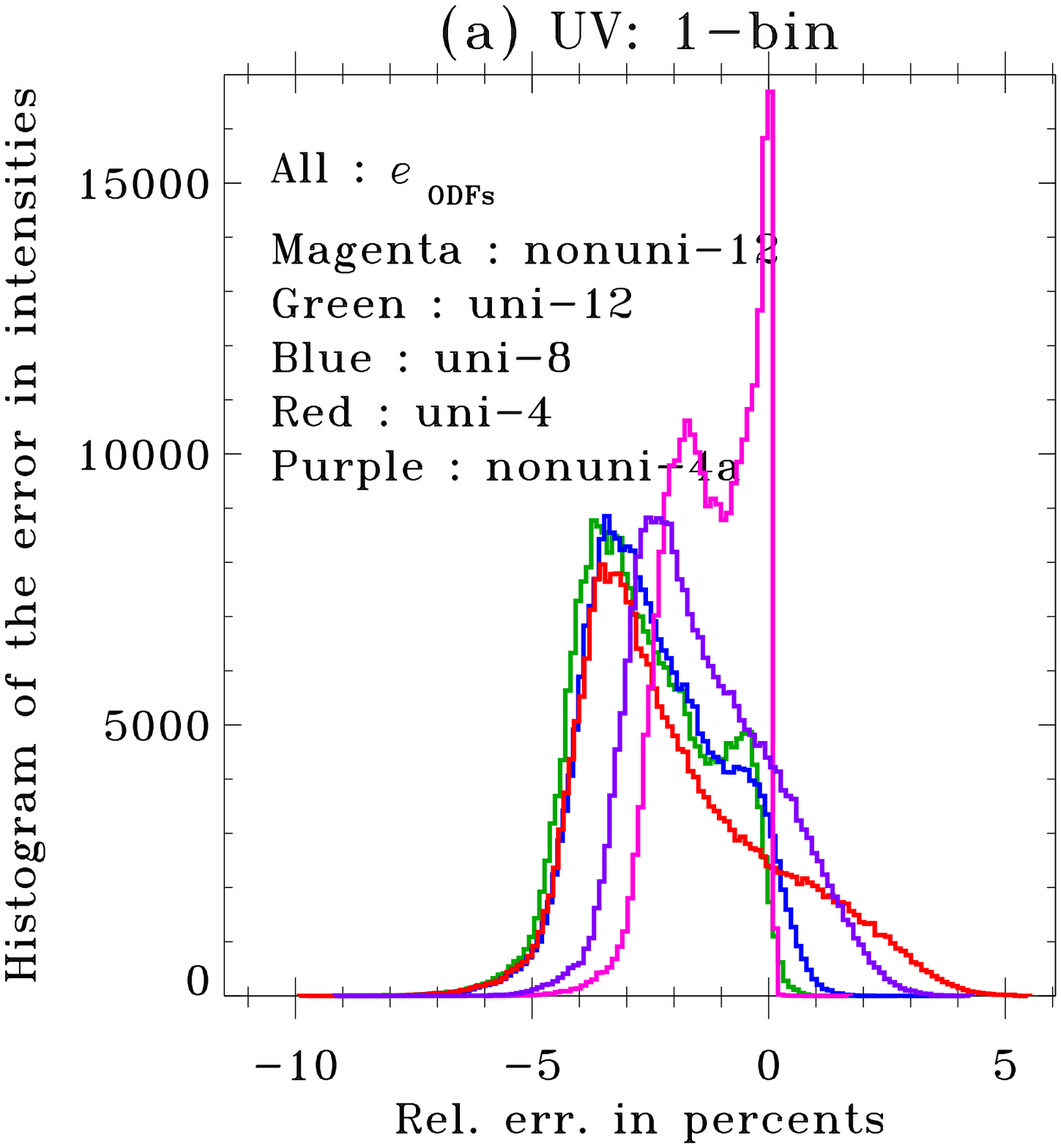}
\includegraphics[scale=0.4]{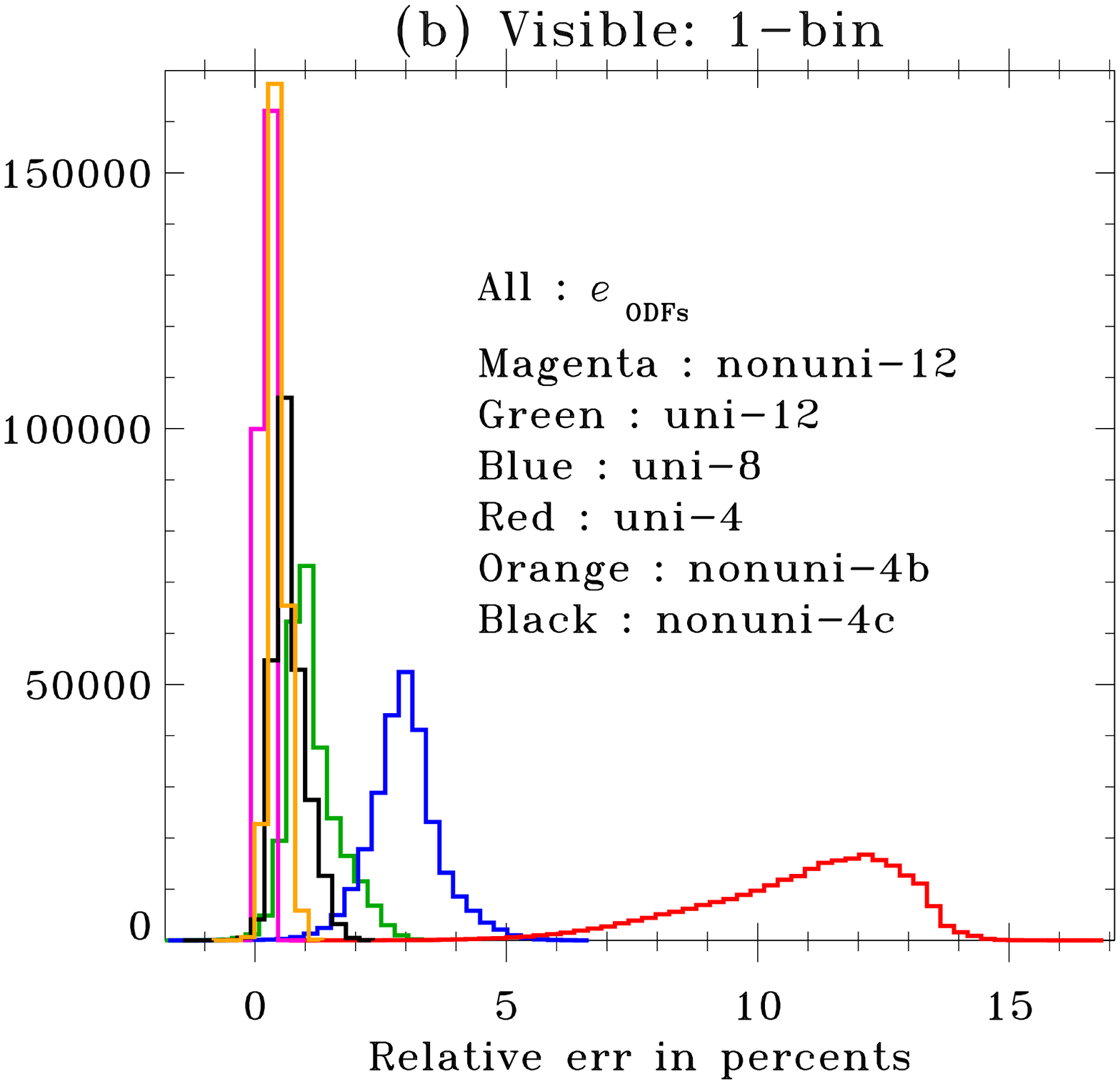}
\caption{Effect of sub-binning when using a single bin to compute the ODFs for the two considered spectral pass-bands. 
Plotted are the histograms of $e_{\textrm {ODFs}}$ for one-bin and various combination of the sub-bin sizes $s_j$ listed in Table~\ref{tab:table2}.  Panels: (a) The UV pass-band. (b) The visible pass-band. 
}
\label{figure4}
\end{figure}

In Figure~\ref{figure4}  we show the distribution of $e_{\textrm {ODFs}}$ for various cases listed in Table~\ref{tab:table2}. 
The different sub-binning configurations we consider are twelve non-uniformly distributed sub-bins as proposed and used by Kurucz, four  non-uniformly distributed sub-bins (see below),  
and twelve, eight and four unifrmly distributed sub-bins. As already discussed in Section~\ref{fodfs-method}, the non-uniform sub-bins should have a coarser division near lower values of opacity and a finer division near the larger values. The configurations using four non-uniform sub-bins are chosen from the study of \citet[][]{2019A&A...627A.157C} (their best configuration at 300 nm, 500 nm and 503 nm). Two of these configurations used in the visible pass-band have a finer division near large opacity values (purple and black lines in Figure~\ref{figure4}b) than the one used in the UV (purple lines in Figure~\ref{figure4}a).

In the UV (Figure~\ref{figure4}a) we observe a two peak distribution of $e_{\textrm {ODFs}}$. 
The peak near zero is highest for Kurucz sub-binning whereas for the uniform sub-binning cases the highest peak appears at larger errors.  In the visible pass-band (Figure~\ref{figure4}b) we see a single peak for all the cases. Here the performance of two  sub-binning configurations from \citet[][]{2019A&A...627A.157C} comes very close to that of the Kurucz sub-binning. 
For both the UV and the visible pass-bands the performance of the ODFs method deteriorates as the number of sub-bins used in the uniform sub-binning cases decreases. This is because it leads to coarser division of the sub-bins near the larger opacity values. Among all the various cases the Kurucz sub-binning case performs the best in both the UV and the visible pass-bands.
 
These examples, computed on a large number of atmospheric structures, re-affirm the findings of \citet[][]{2019A&A...627A.157C} that the accuracy of the ODFs is sensitive to the non-uniformity of the sub-bins, in particular near larger opacity values. 
\subsection{Performance of FODFs}
\label{result:fodfs_performance}
In this section we study the perfomance of the FODFs method for computing intensities passing through non-rectangular filters. For this purpose we use the  filter profiles SuFI and SuFIV (see their description at the end of Section~\ref{1.5D-RT}).

\begin{figure}
\includegraphics[scale=0.4]{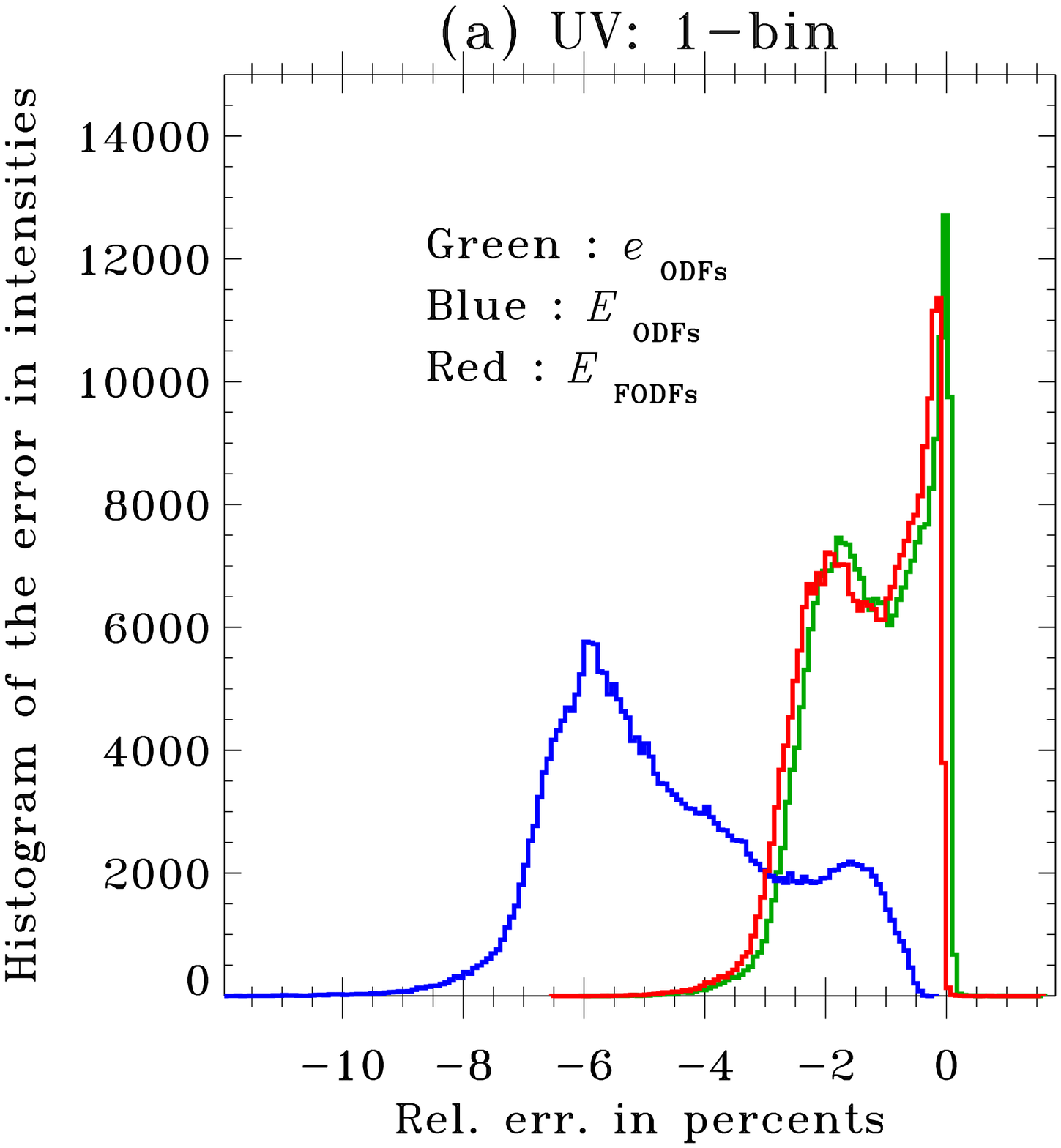}
\includegraphics[scale=0.4]{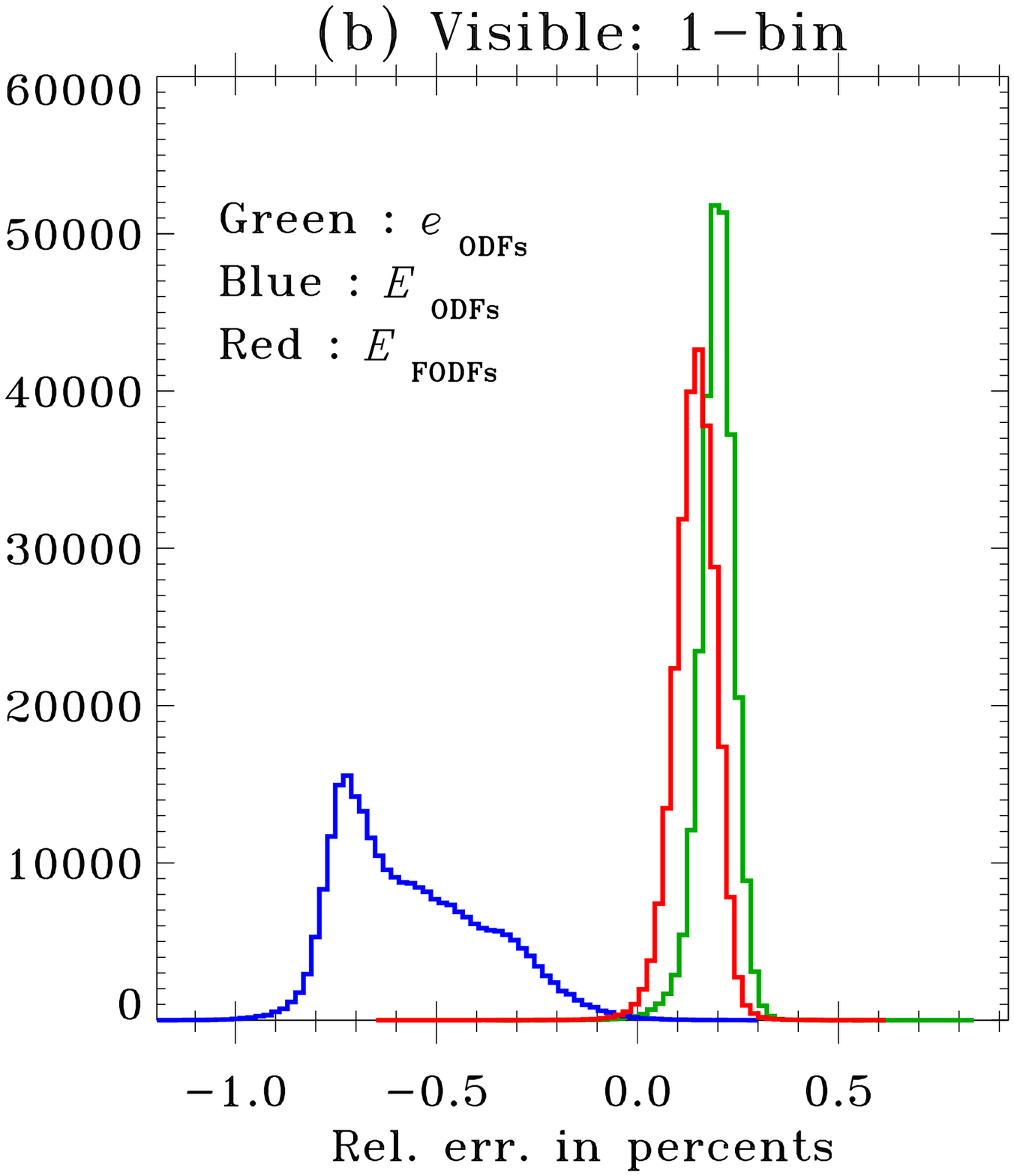}
\caption{Performance of FODFs. Plotted are the histograms of $e_{\textrm {ODFs}}$, $E_{\textrm {ODFs}}$ and $E_{\textrm {FODFs}}$ for the  one-bin set-up with the Kurucz sub-binning. Panels:  (a) The SuFI filter. (b) The SuFIV filter.  
}
\label{figure5}
\end{figure}

In Figure~\ref{figure5} we show the distributions of $e_{\textrm {ODFs}}$, $E_{\textrm {ODFs}}$ and 
$E_{\textrm {FODFs}}$ in the UV and the visible pass-bands. The $e_{\textrm {ODFs}}$ and $E_{\textrm {FODFs}}$ distributions in general look very similar. Their respective mean values are also similar, namely, in the UV pass-band they are -1.2 \% and -1.4 \% while in the visible pass-band they are 0.19 \% and 0.14 \%. 
This implies that the performance of the FODFs method for intensities passing through non-rectangular filters is as good as that of the traditional ODFs method for intensity passing through rectangular filters. This is not surprising since the entire idea of the FODFs method is to mimic the performance of the ODFs method for non-rectangular filters. Indeed, the FODFs method essentially boils down to the ODFs method but performed on a modified frequency grid (see Equation~\ref{eq:tilda} and discussion in Section~\ref{fodfs-method}). The mean of the $E_{\textrm {ODFs}}$ distribution in the UV and the visible pass-bands are -4.6\% and -0.56\% respectively. It is clear from Figure~\ref{figure5} (blue and red curve) that the FODFs method performs better than the ODFs method for computation of the intensities passing through the non-rectangular filters.

\begin{figure}
\includegraphics[scale=0.4]{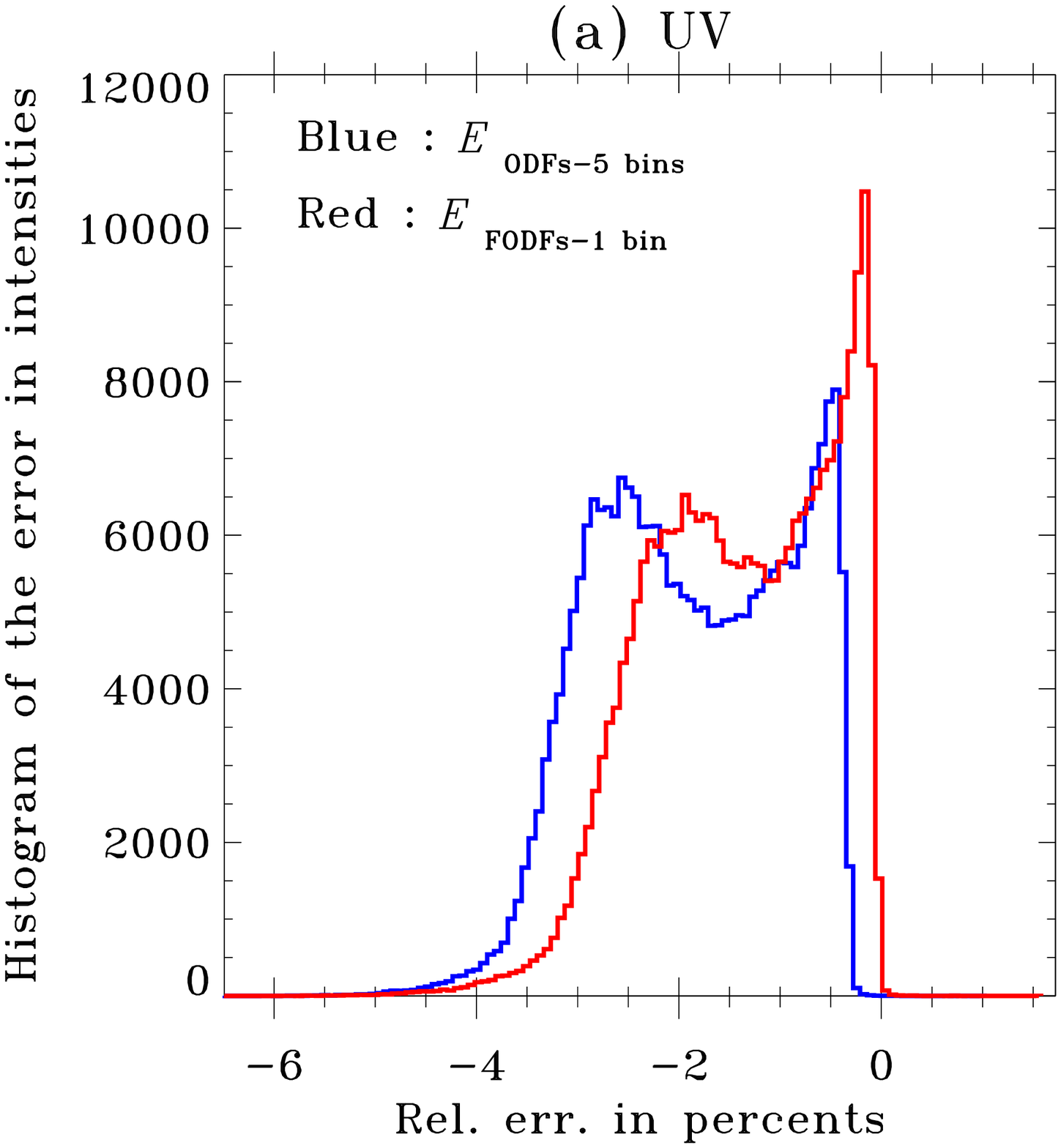}
\includegraphics[scale=0.4]{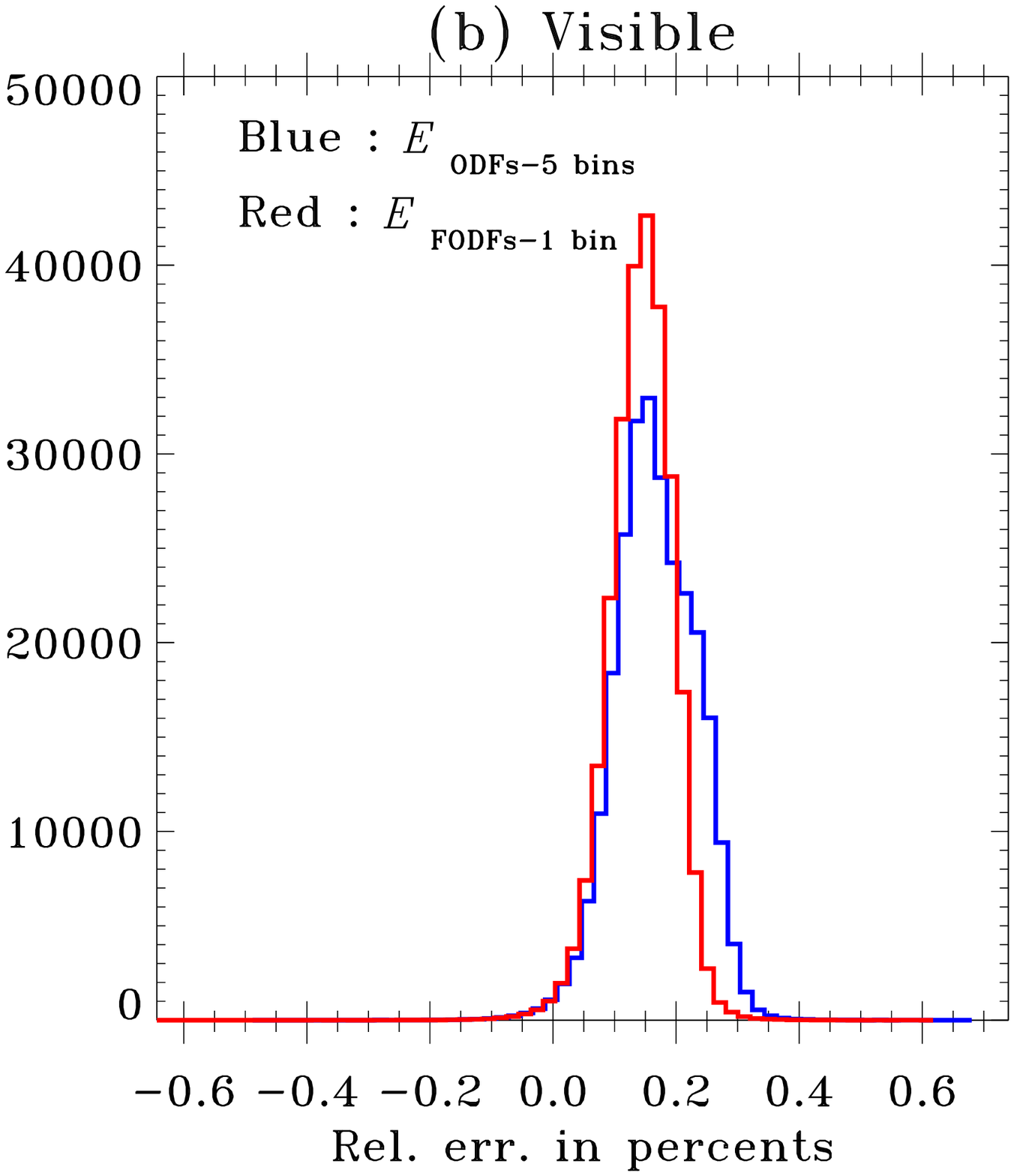}
\caption{Performance of FODFs. Plotted are the histograms of $E_{\textrm {ODFs}}$ for the five-bins set-up and $E_{\textrm {FODFs}}$ for the one-bin set-up with the Kurucz sub-binning. Panels:  (a) The SuFI filter. (b) The SuFIV filter.  
}
\label{figure55}
\end{figure}

One way to improve the performance of the ODFs method for calculating intensities through the non-rectangular filters is to increase the number of bins in the pass-band of the filter. For a sufficiently large number of bins, one can neglect the change of the transmission function within the bin and the $F_{\textrm {ODFs}}$ becomes a good approximation of the $F_{\textrm{HR}}$. An example of the improved performance of ODFs is demonstrated in Figure~\ref{figure55} where we show  $E_{\textrm {ODFs}}$ computed using five bins and compare them with 
$E_{\textrm {FODFs}}$ computed using one bin. 
Now the mean of the $E_{\textrm {ODFs}}$ ($E_{\textrm {FODFs}}$) distribution in the SuFI and the SuFIV filters are -1.9\% (-1.4\%) and -0.16\% (-0.14\%) respectively. This shows that the performance of ODFs, indeed, improves (compare blue lines in Figures \ref{figure5} and \ref{figure55}) and almost approaches the performance of FODFs (although is still a bit worse). 
However, this implies that RT calculations have to be performed on five times larger number of frequencies which increases the computational costs by five times. The FODFs method is free from such a shortcoming and thus, is five times faster than the traditional ODFs method.

\begin{figure}
\includegraphics[scale=0.55]{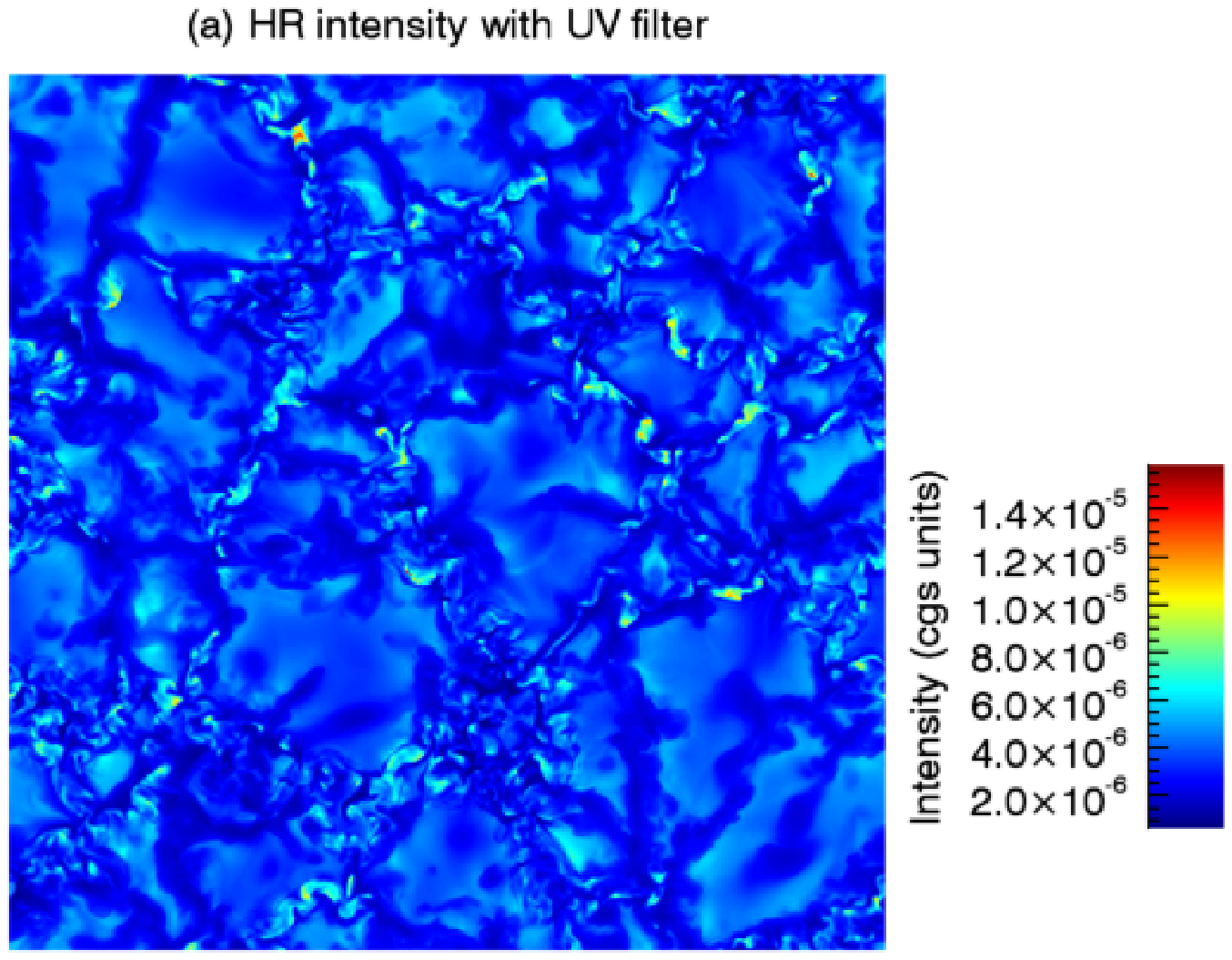}
\includegraphics[scale=0.55]{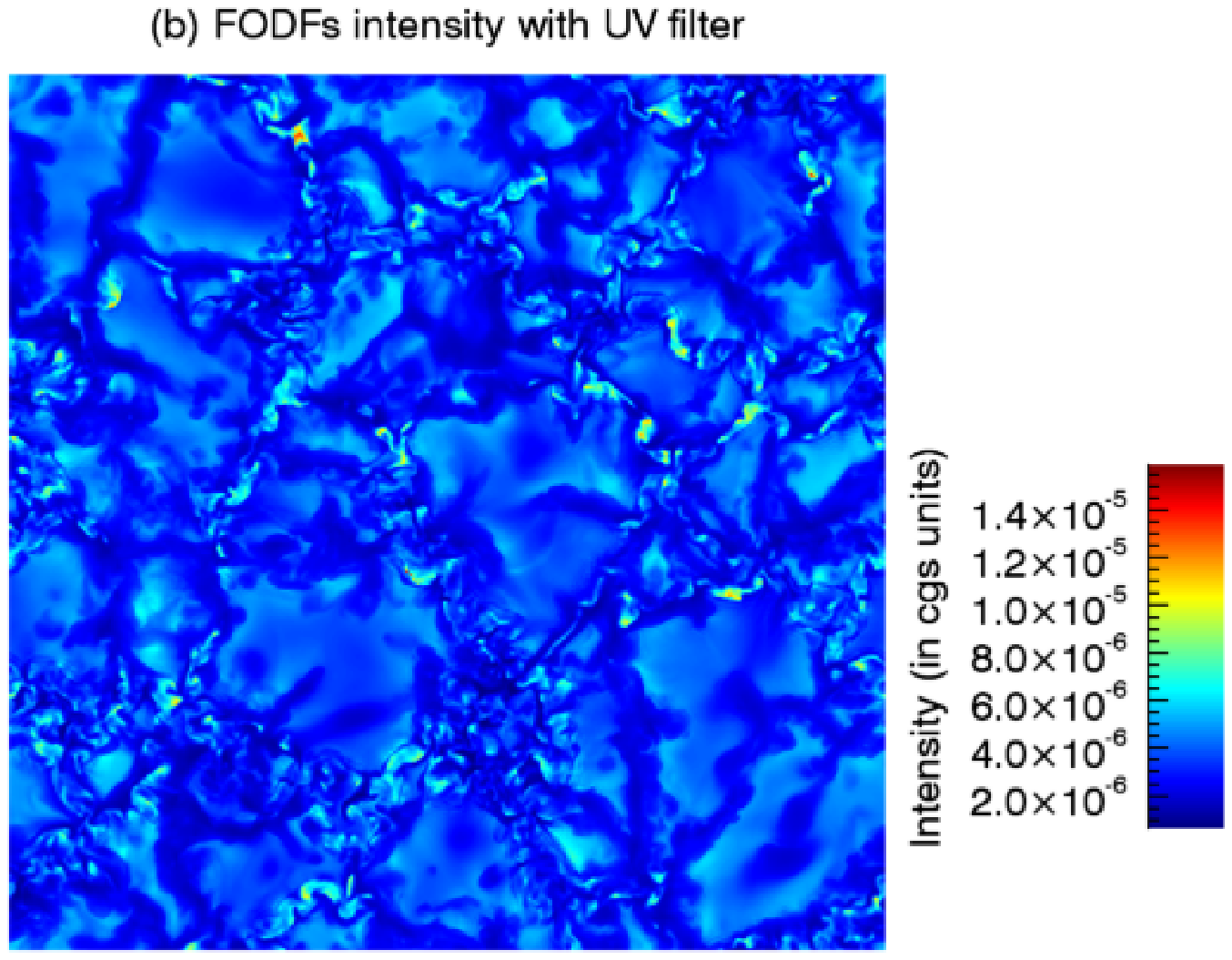}
\caption{Images of the spectral intensities seen through SuFI filter for the one-bin set-up with the Kurucz sub-binning. Panels: (a) High-spectral resolution intensity. (b) Intensity computed using the FODFs method.  
}
\label{map-flux-HR-fodf-sufi}
\end{figure}

\begin{figure}
\includegraphics[scale=0.55]{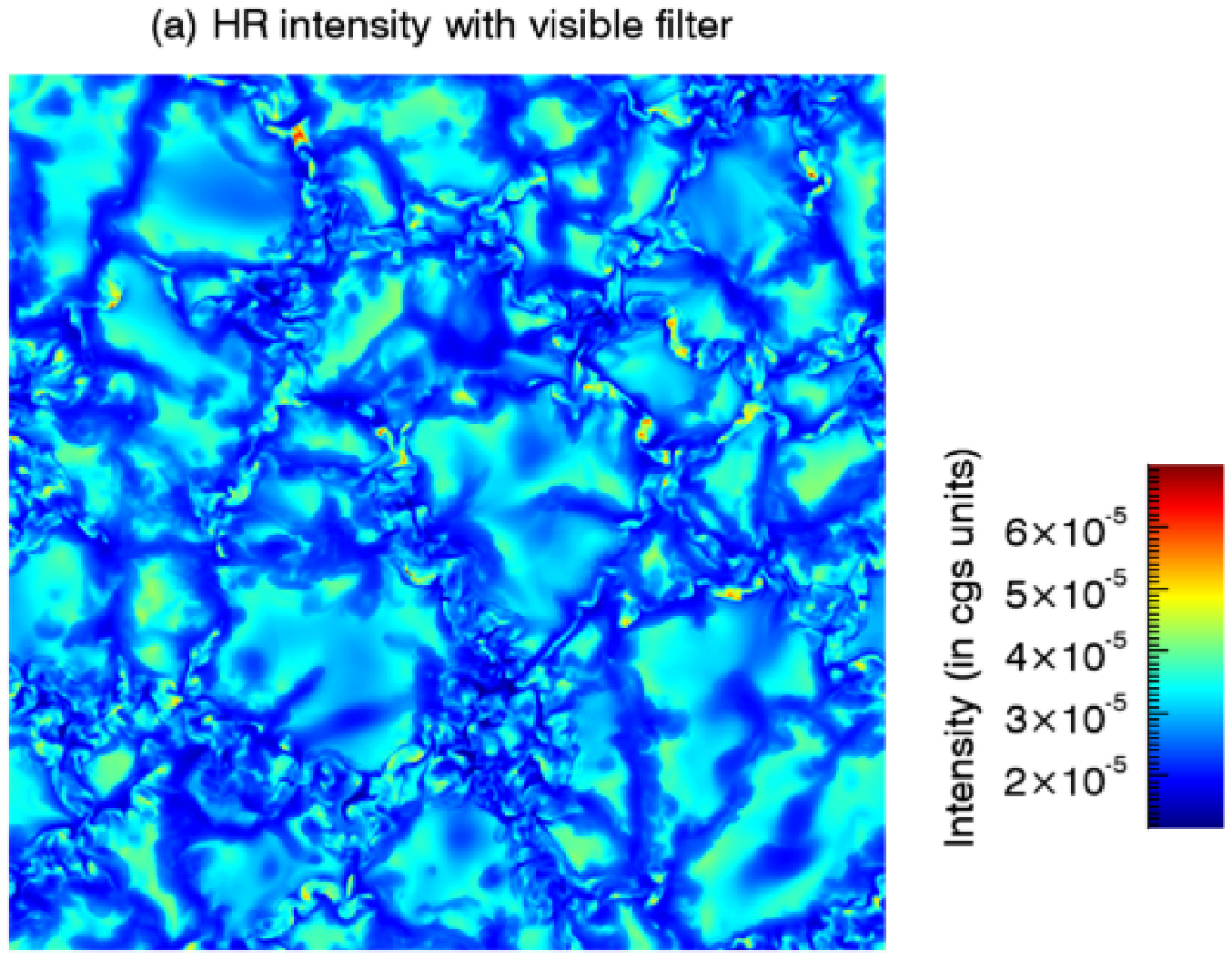}
\includegraphics[scale=0.55]{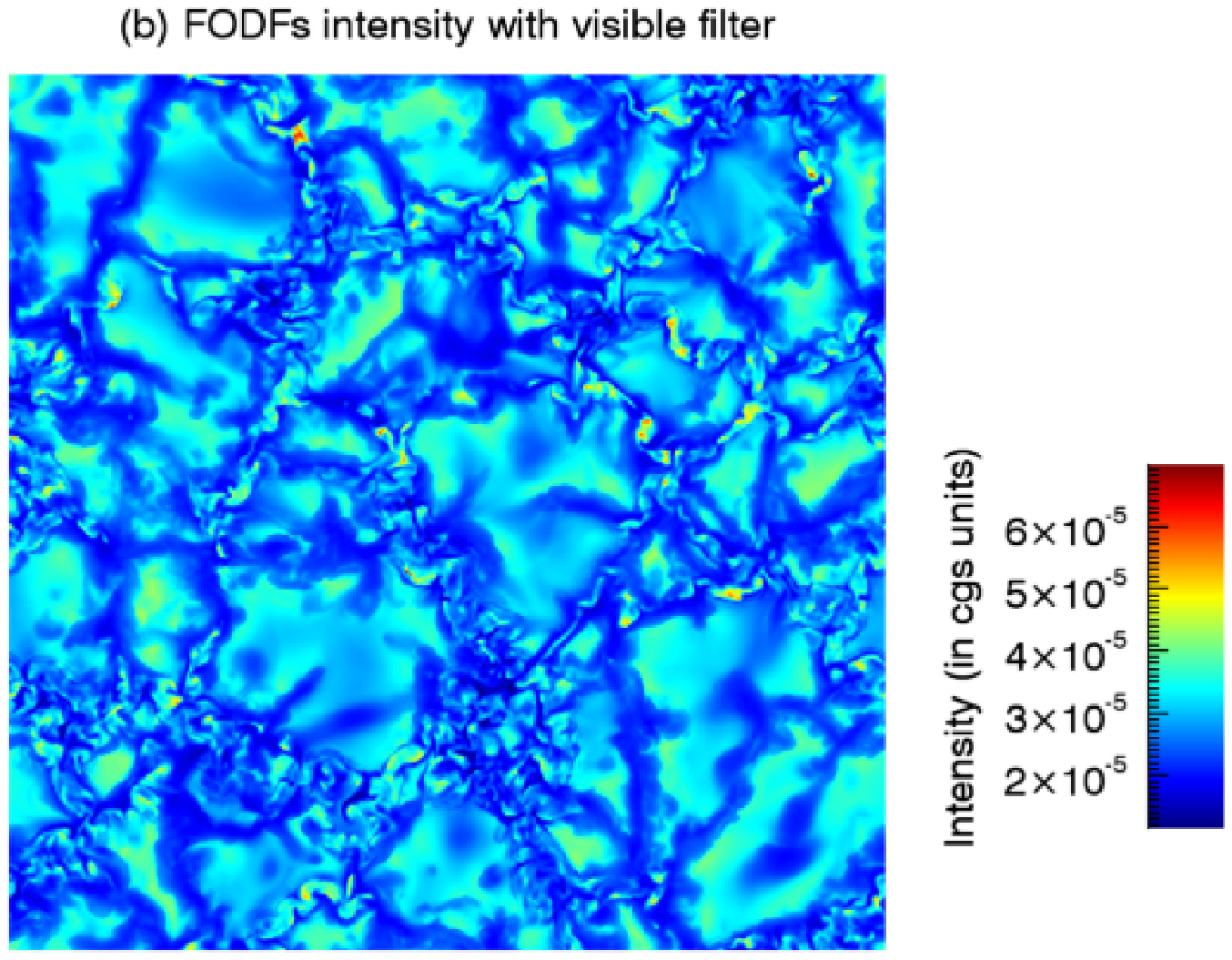}
\caption{Images of the spectral intensity seen through SuFIV filter for the one-bin set-up with the Kurucz sub-binning. Panels: (a) High-spectral resolution intensity. (b) Intensity computed using the FODFs method.  
}
\label{map-flux-HR-fodf-visible}
\end{figure}

In Figures~\ref{map-flux-HR-fodf-sufi} 
and \ref{map-flux-HR-fodf-visible} we show the images of the emergent spectral intensities on the entire 300 G cube in the UV and the visible pass-bands respectively. In each of these figures shown are the intensities computed using high-spectral resolution case and the FODFs method. These figures show that the FODFs method is accurate enough to reproduce the important spatial structuring reflected in the images of the intensities. 

\begin{figure}
\includegraphics[scale=0.55]{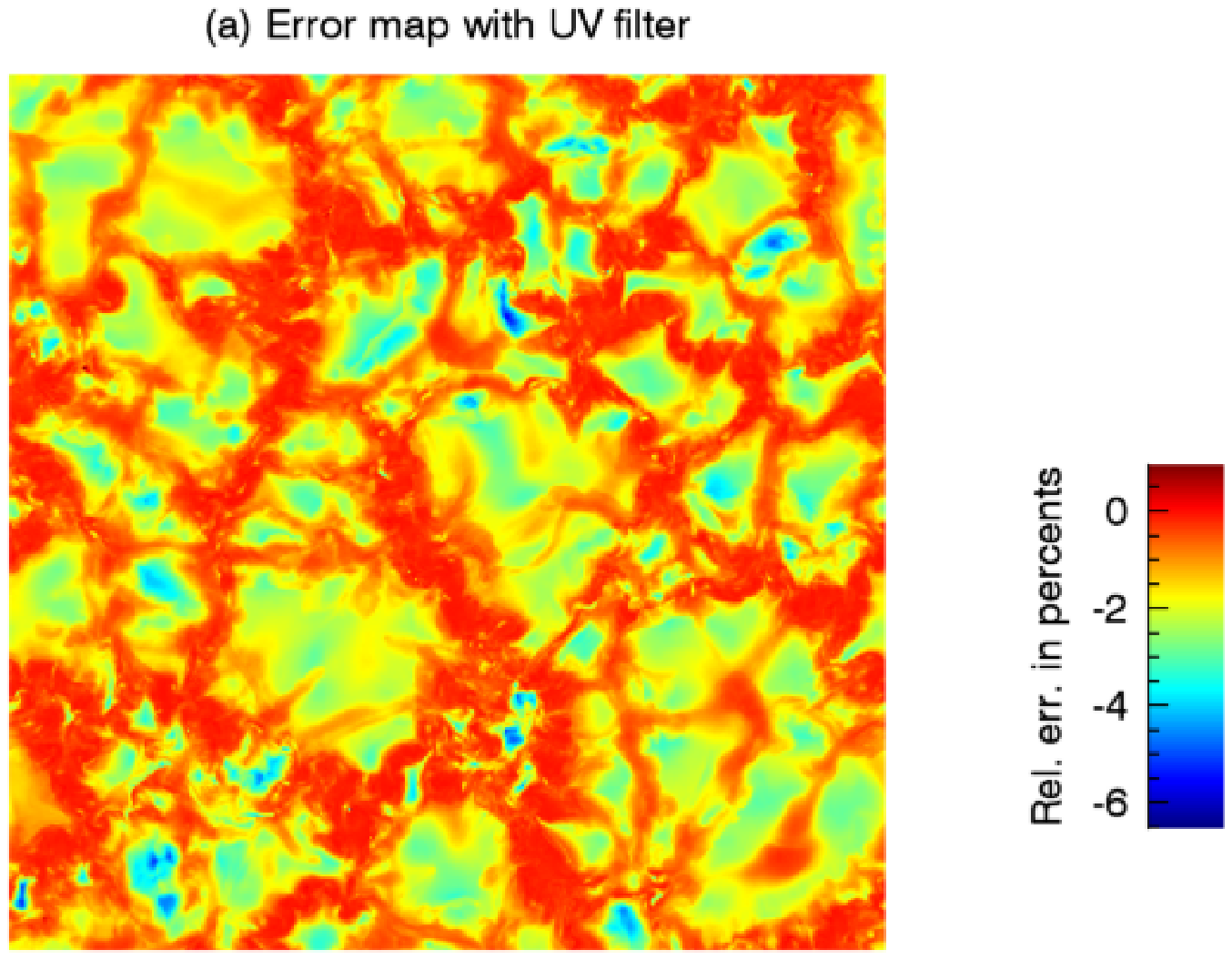}
\includegraphics[scale=0.55]{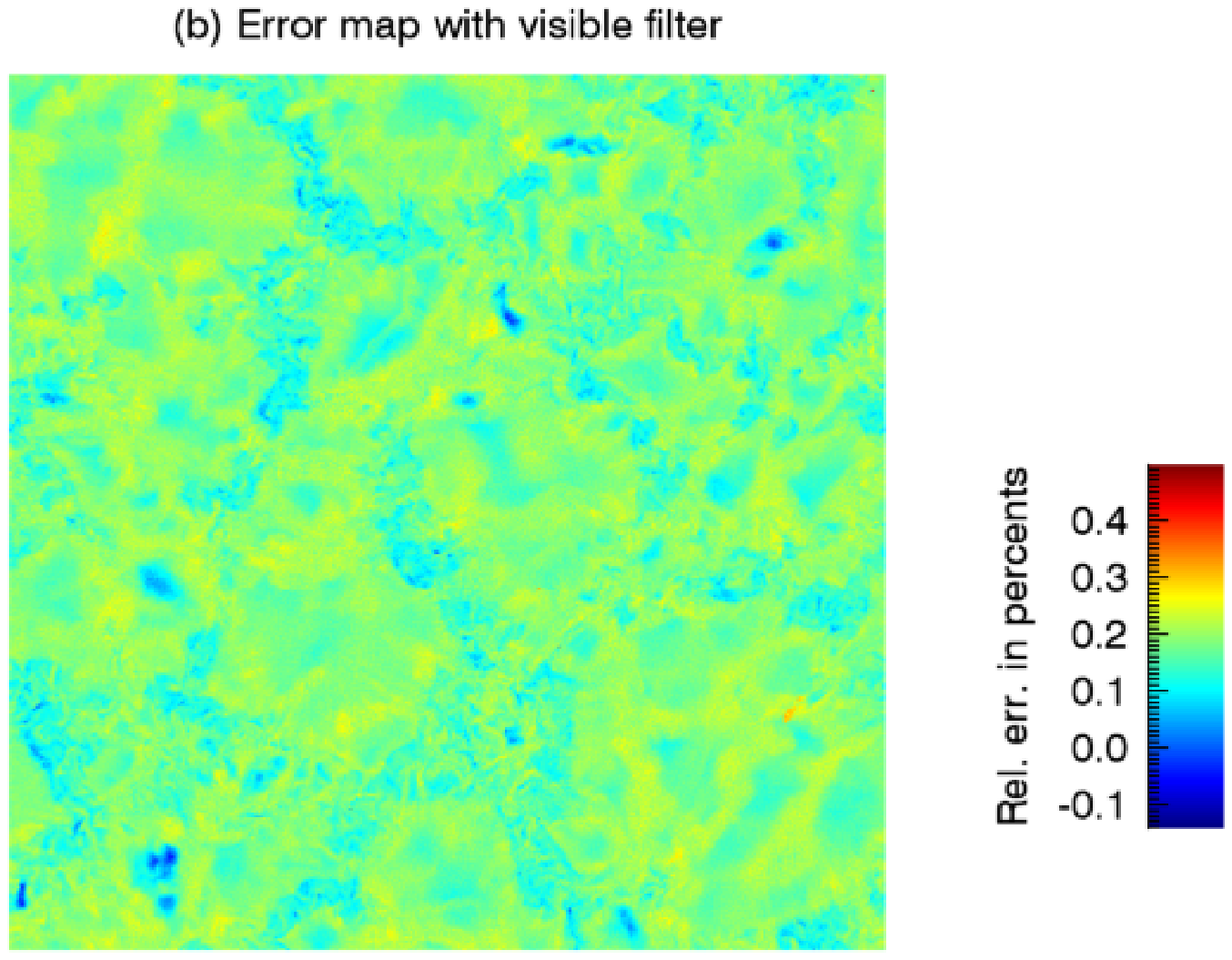}
\caption{Images of the $E_{\textrm {FODFs}}$ with intensities computed using one-bin and the Kurucz sub-binning set-up. Panels: (a) For SuFI filter. (b) For SuFIV filter.  
}
\label{map-err-HR-fodf}
\end{figure}

To visualize the error distribution 
we show the images of the $E_{\textrm {FODFs}}$ in Figure~\ref{map-err-HR-fodf} for the UV and the visible pass-bands. A comparison of the images of intensities in Figures~\ref{map-flux-HR-fodf-sufi} 
and \ref{map-flux-HR-fodf-visible} with the error maps in Figure~\ref{map-err-HR-fodf} indicates that they have a striking structural correlation in the UV while in the visible pass-band the correlation is weak.
In the UV pass-band the magnitude of the errors is higher in the brighter granular structures and lower in darker inter-granular structures, while in the visible pass-band they are nearly uniform.  In particular in the UV pass-band the individual errors can be as high as -6.5\%. 
We note that a detailed study of the source of these individual large errors is beyond the scope of this paper.
However, as already discussed above the mean errors are much smaller. This is because the large individual errors occur only at certain grid points and do not dominate the distribution.

{\bf{
\begin{figure}
\includegraphics[scale=0.4]{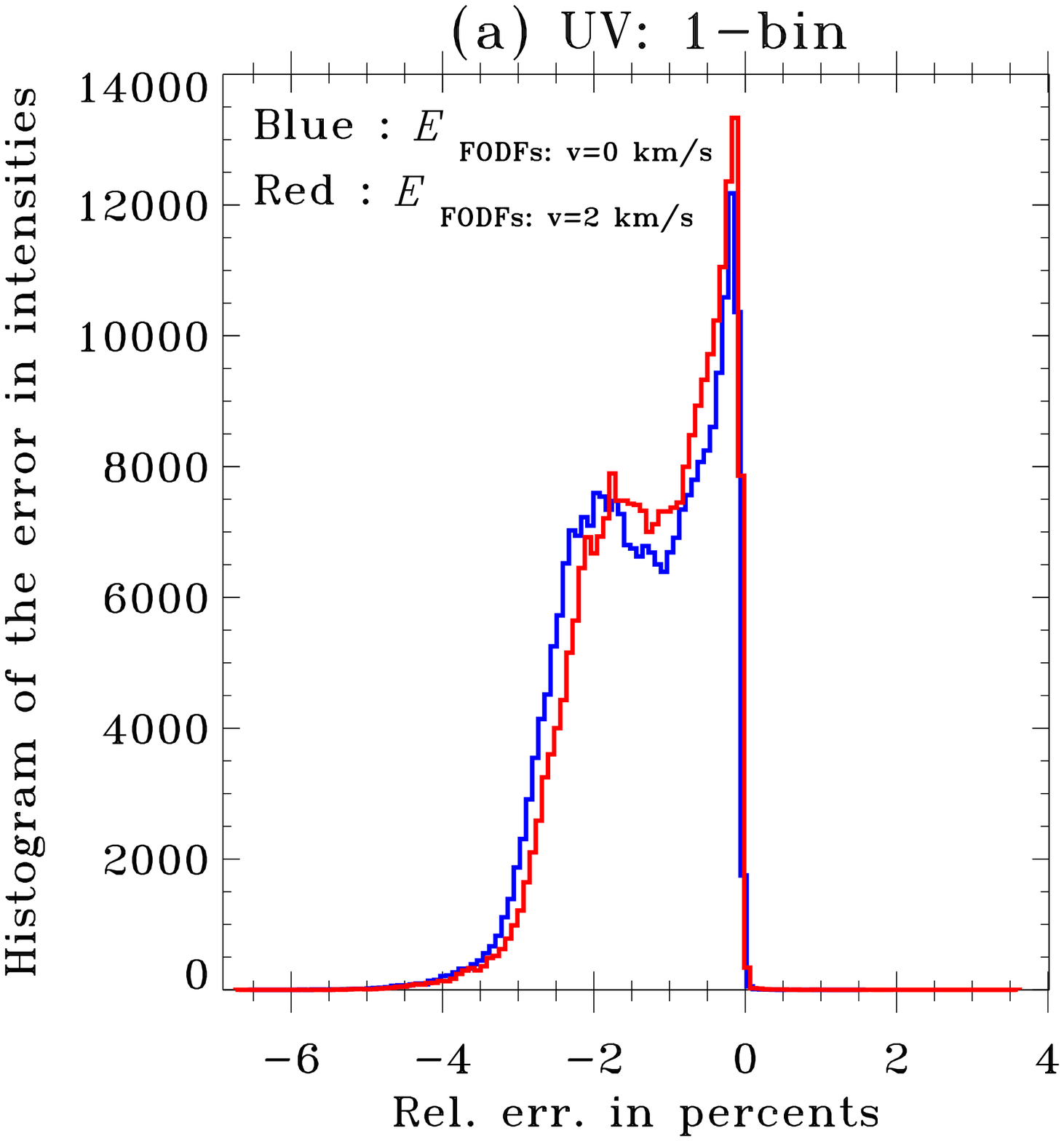}
\includegraphics[scale=0.4]{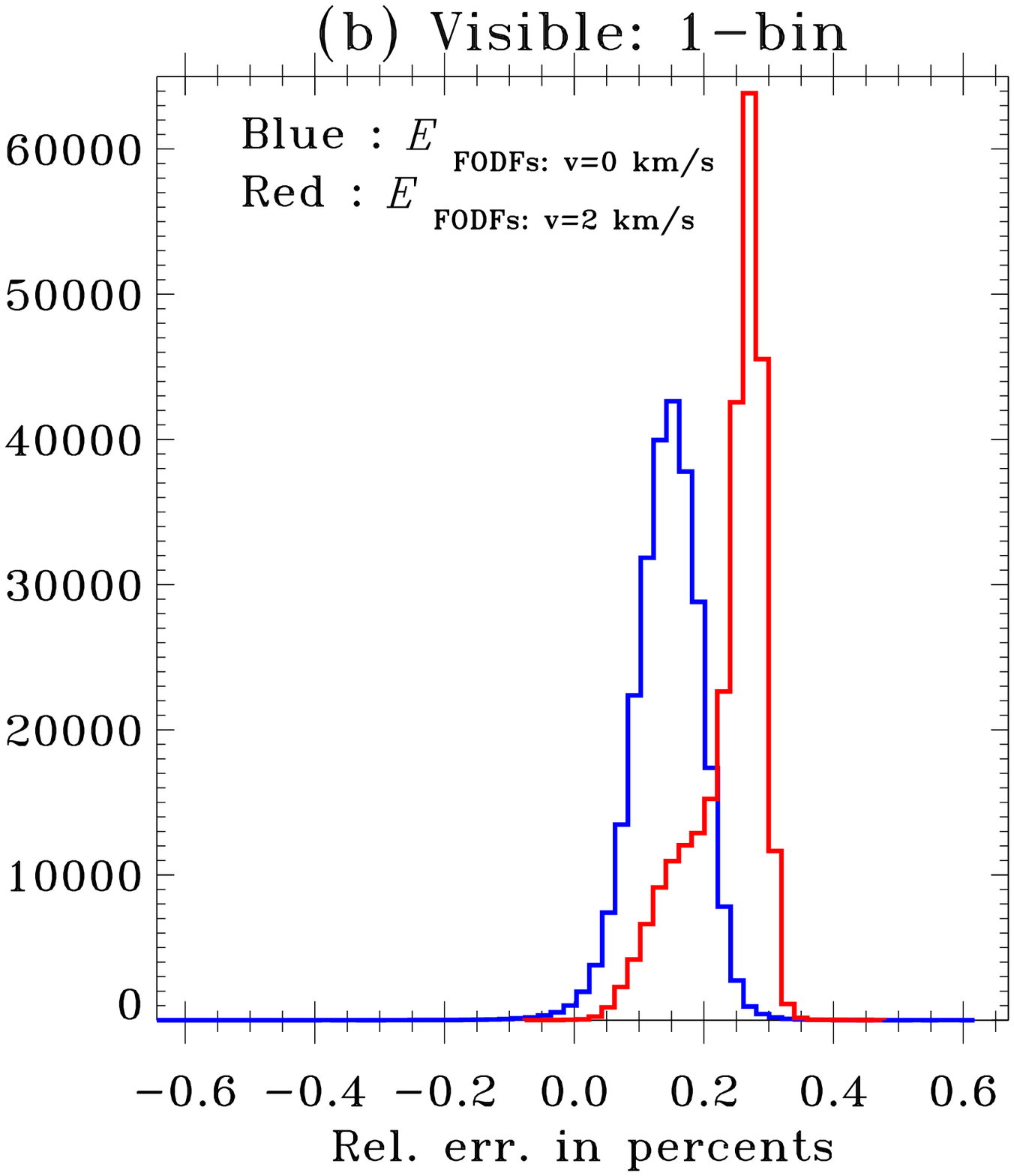}
\caption{Effect of micro-turbulent velocity ${\rm{v}}$. Plotted are the histograms of $E_{\textrm {FODFs}}$ for the one-bin set-up and the Kurucz sub-binning with ${\rm{v}}$ values 0 and 2 km/s.  Panels:  (a) The SuFI filter. (b) The SuFIV filter.  
}
\label{figure5AB}
\end{figure}

Since the main goal of the paper is to validate the method of FODFs by comparing with the high resolution solutions and the solutions using ODFs, for all the studies until now we simply chose the micro-turbulent velocity value to be 0 km/s as an example. In Figure~\ref{figure5AB} we illustrate the effect of micro-turbulent velocity ${\rm{v}}$ on the distributions of $E_{\textrm {FODFs}}$ in the UV and the visible pass-bands. We compare the $E_{\textrm {FODFs}}$ for ${\rm{v}}$ values of 0 km/s and 2 km/s. The respective mean values of $E_{\textrm {FODFs}}$ for ${\rm{v}}=$ 0 km/s and 2 km/s are, -1.4 \% and -1.25 \%  in the UV pass-band while in the visible pass-band they are 0.14 \% and 0.23 \%. Thus the use of non-zero micro-turbulent velocity values leads to slightly smaller errors in the UV and slightly larger errors in the visible pass-bands. 
}}

\subsection{Center-to-limb Variation}
\label{clv}
\begin{figure}
\includegraphics[scale=0.4]{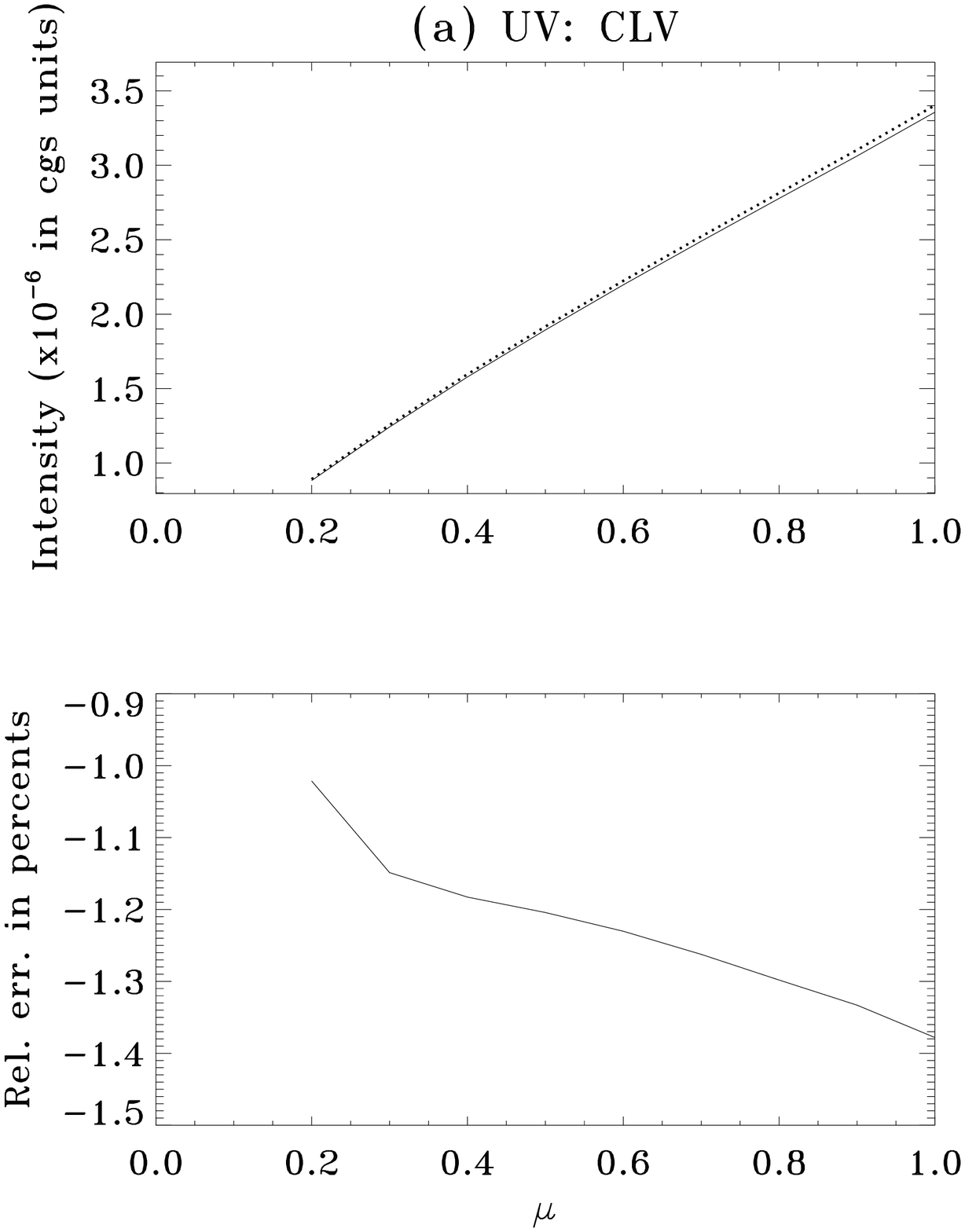}
\includegraphics[scale=0.4]{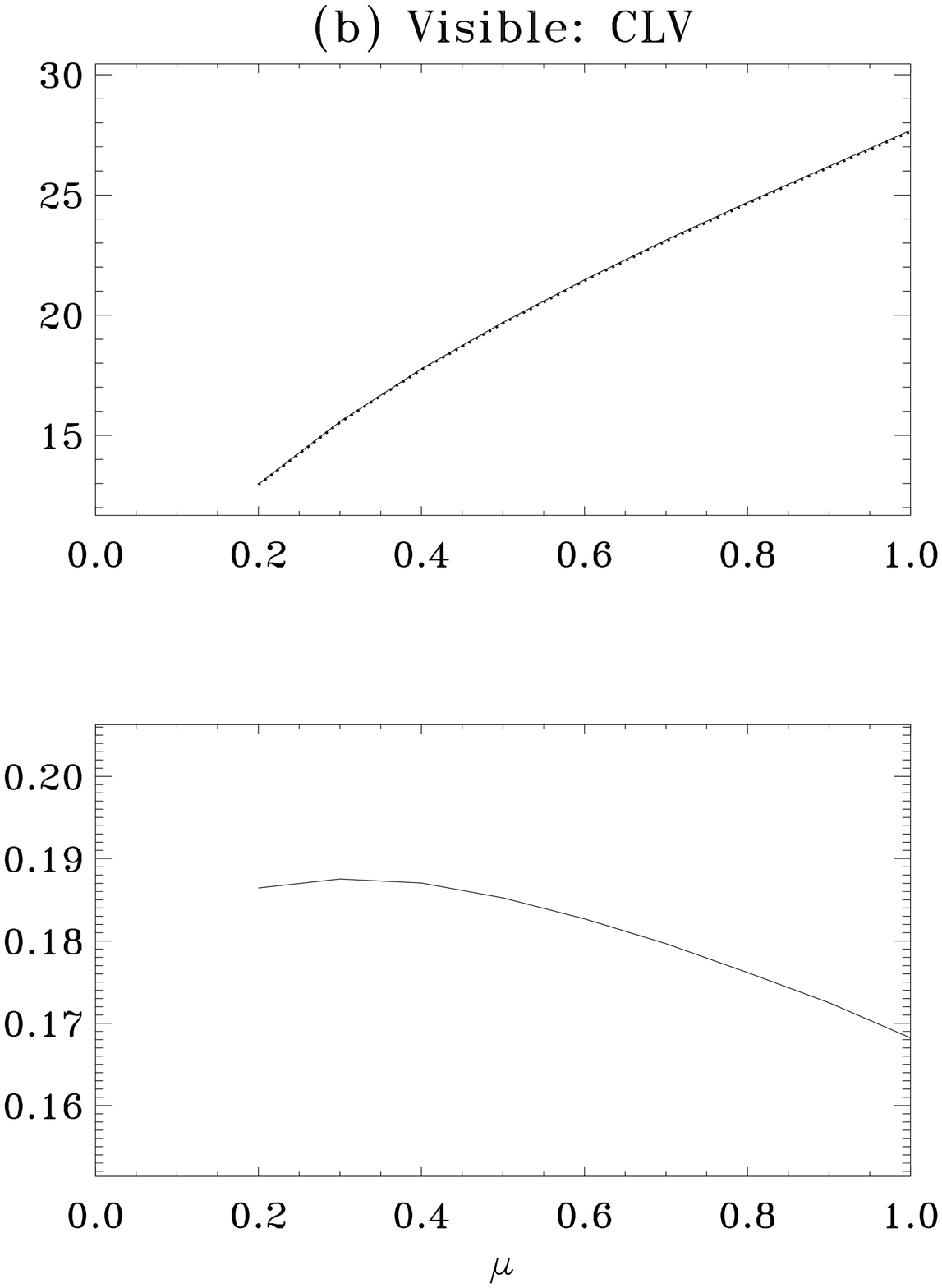}
\caption{In the top panels we plot the center-to-limb variation of the spatially averaged emergent spectral intensities computed using high-spectral resolution (solid lines) and the FODFs method (dotted lines) through non-rectangular filters, while in the bottom panels the corresponding $E_{\textrm {FODFs}}$ are plotted. These calculations correspond to one-bin set-up with Kurucz sub-binning.  Panels: (a) The SuFI filter. (b) The SuFIV filter.  
}
\label{figure9}
\end{figure}

In the previous sections we only considered the radiation along vertical rays. In this section we consider rays at nine inclinations equally spaced in $\mu$ between $\mu=0.2$ and 1, where $\mu=\cos \theta$ and $\theta$ is the angle between the ray and the normal vector. For each of the rays we spatially average the emergent spectral intensities. We show this center-to-limb variation of the spatially averaged intensities passing through the SuFI and the SuFIV filters in the top panels of Figure~\ref{figure9}. The two curves in each of these panels are computed with high-spectral resolution and with the FODFs method.  In the bottom panels we plot the $E_{\textrm {FODFs}}$ between the two intensities plotted in the top panels. 

Figure~\ref{figure9} shows that the absolute relative errors of intensity values are below 1.4\% in the SuFI filter and below 0.19\% in the SuFIV filter. An accurate calculation of the center-to-limb intensity variations is essential for the characterization of exoplanets from transit light curves. We note, that the transit light curves do not depend on the offset in the intensity values (i.e., error independent of $\mu$) so that only the change of the error with $\mu$ has an effect on exoplanet characterization. One can see that the errors brought about by the usage of FODFs show a rather weak dependence on the $\mu$ value:  in the SuFI filter the absolute relative error increases towards disk-center, while in the SuFIV filter the error decreases towards the disk-center. The change in the error in Figure~\ref{figure9} between $\mu=0.2$ and  $\mu=1$ is only 0.36 \% in the SuFI and 0.018 \% in the SuFIV filter. 

\section{Conclusions}
\label{conclusions}
In this paper we have tested the performance of the ODFs method for calculating spectral intensities emerging from 3D cubes representing solar atmospheres.  We  show that ODFs return accurate values of intensities and their centre-to-limb variations. We also demonstrate that sub-bin configurations which were found to be good for simple 1D models also work well for the 3D case.

Furthermore, we have generalised the ODFs concept  for calculating  intensities in narrow-band non-rectangular filters, introducing filter-ODFs or FODFs. We recall here that by narrow band filters we mean the filters narrow enough that neither the continuum opacity nor the Planck function changes noticeably within it. Further we recall that the rectangular filter refers to the one which is zero outside of the given spectral pass-band and has a constant value within this pass-band.  
We have implemented this method in the MPS-ATLAS code and tested it rigorously. 
By formulation the FODFs method is identical to the traditional ODFs method but with a change-of-variable for integration.

The main advantage of the FODFs method is that the intensities in the narrow band non-rectangular filters can be computed in just one wavelength bin (or interval) with the intrinsic accuracy of the traditional ODFs method in rectangular filters. We analyzed our FODFs method using two non-rectangular filters (i.e., one in the UV and the other in the visible pass-band). In both the cases the FODFs method does an excellent job. Due to its advantage in terms of its speed and accuracy, this method finds its applicability in studies of stellar variability and exo-planet detection and characterization, population synthesis of stars in galaxies.

As next steps we would like to extend our studies to broader spectral pass-bands and test the performance of the method for stars with various effective temperatures.  In particular, we will find the optimal setups of FODFs for calculating intensities in pass-bands routinely used in the ground based photometry (e.g. in Str\"omgen and Johnson-Morgan pass-band) as well as in the  space-borne photometry (e.g., in pass-bands of {\it {Kepler}}, TESS, CHEOPS, and Gaia telescopes).
We expect that our method will be useful for a broad range of applications demanding calculations of intensity in spectral pass-bands, e.g. stellar colours, photometric variability of the Sun and stars, stellar limb darkening in pass-bands of transit-photometry telescopes (which is crucial for the determination of exoplanet radii), etc. At the same time our approach is not suitable for the  synthesis of the individual line profiles.

\acknowledgements{The research leading to this paper has received funding from the European Research Council under the European Union’s Horizon 2020 research and innovation program (grant agreement No. 715947). It also got financial support from the BK21 plus program through the National Research Foundation (NRF) funded by the Ministry of Education of Korea. The computational resources were provided by the German Data Center for SDO through
grant 50OL1701 from the German Aerospace Center (DLR).
This work was partially supported from ERC Synergy Grant WHOLE SUN 810218.}

\bibliography{ms}{}
\bibliographystyle{aasjournal}

\end{document}